\documentclass[a4paper,twocolumn,11pt,accepted=2021-02-03]{quantumarticle}
\pdfoutput=1
\usepackage[numbers,sort&compress]{natbib}

\usepackage[utf8]{inputenc}
\usepackage{xspace}
\usepackage[french,english]{babel}
\usepackage[T1]{fontenc}
\usepackage[margin=28mm,includeheadfoot,bindingoffset=0mm]{geometry}
\usepackage{graphics,graphicx,keyval,subfigure}
\usepackage[normalem]{ulem}
\usepackage{fancyhdr}
\usepackage{stmaryrd}
\usepackage[x11names,svgnames]{xcolor}
\usepackage{amsmath, mathtools}
\usepackage{amsfonts,bm,bbm}
\usepackage{upgreek}
\usepackage{etoolbox}
\usepackage{calc}
\usepackage{datetime}
\usepackage[squaren,Gray]{SIunits}
\usepackage{cancel}
\usepackage{amsmath,amsfonts}
\usepackage{amssymb}
\usepackage{mathtools}
\usepackage{slantsc}
\usepackage{hyperref}
\usepackage[all]{hypcap}
\usepackage{tikz}
\usepackage{lipsum}
\usepackage{blindtext}
\usepackage{perpage}
\usepackage{physics}
\usepackage{textcomp}
\usepackage{bbold}
\usepackage{multirow}
\usepackage{enumerate}

\newcommand\st{\textsuperscript{st}\xspace}
\newcommand\nd{\textsuperscript{nd}\xspace}

\newcommand\nth{\textsuperscript{th}\xspace}

\newcommand{\tens}[1]{
    \mathbin{\mathop{\otimes}\limits_{#1}}
  }

\begin{document}

\title{Resource requirements for efficient quantum communication using all-photonic graph states generated from a few matter qubits}
\author{Paul Hilaire}
\orcid{0000-0002-7144-6953}
\affiliation{Department of Physics, Virginia Tech, Blacksburg, Virginia 24061, USA}
\email{paulhilaire@vt.edu}
\author{Edwin Barnes}
\affiliation{Department of Physics, Virginia Tech, Blacksburg, Virginia 24061, USA}
\orcid{0000-0003-1666-9385}
\author{Sophia E. Economou}
\affiliation{Department of Physics, Virginia Tech, Blacksburg, Virginia 24061, USA}
\orcid{0000-0002-1939-5589}

\begin{abstract}
    Quantum communication technologies show great promise for applications ranging from the secure transmission of secret messages to distributed quantum computing. Due to fiber losses, long-distance quantum communication requires the use of quantum repeaters, for which there exist quantum memory-based schemes and all-photonic schemes. While all-photonic approaches based on graph states generated from linear optics avoid coherence time issues associated with memories, they outperform repeater-less protocols only at the expense of a prohibitively large overhead in resources. Here, we consider using matter qubits to produce the photonic graph states and analyze in detail the trade-off between resources and performance, as characterized by the achievable secret key rate per matter qubit. We show that fast two-qubit entangling gates between matter qubits and high photon collection and detection efficiencies are the main ingredients needed for the all-photonic protocol to outperform both repeater-less and memory-based schemes.
\end{abstract}
\maketitle

The ability to share entangled states over long distances is a major milestone for the realization of a fully-functional quantum internet~\cite{Kimble2008, Wehner2018}.
Beyond secure communications via quantum key distribution (QKD)~\cite{Bennett1984, Bennett1992,Jennewein2000}, the implementation of such a quantum internet would also have various applications ranging from distributed quantum computing~\cite{Broadbent2009}, secure access to a remote quantum computer~\cite{Grover1997, Nickerson2014}, accurate clock synchronization~\cite{Komar2014}, and improved telescope observations~\cite{Gottesman2012}.

However, enabling world-wide quantum communication requires addressing the major problem of photonic losses, which significantly reduces the range of quantum information transfer.
Even though direct amplification of a quantum state is made impossible due to the no-cloning theorem~\cite{Wootters1982, Dieks1982}, this exponential photon loss can still be overcome through the realization of quantum repeaters (QR)~\cite{Briegel1998, Dur1999}.

\begin{figure*}
    \centering
    \includegraphics[width=17cm]{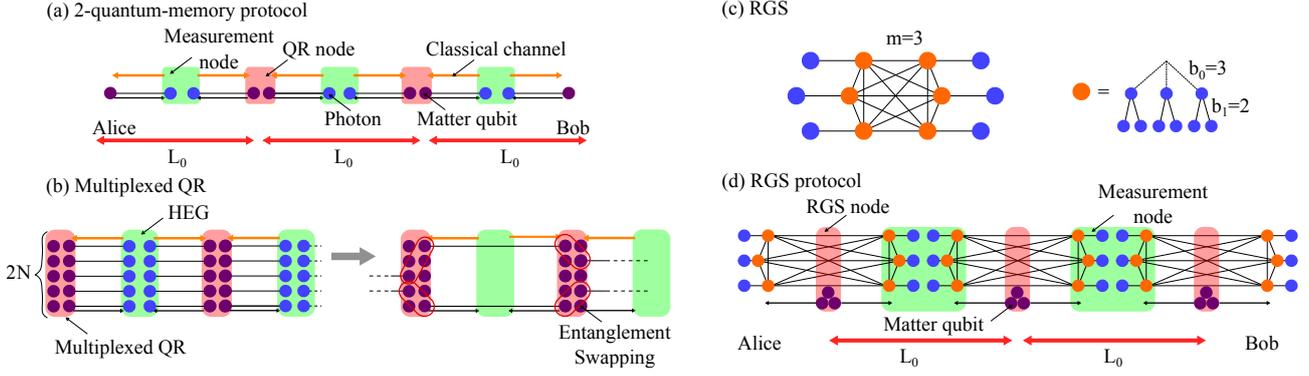}
    \caption{(a) 2-quantum-memory QR protocol. (b) Multiplexed quantum repeater composed of $2N$ memories per repeater with heralded entanglement generation (HEG) at measurement nodes. (c) Left side: Example of a repeater graph state. Vertices represent qubits (blue: physical qubits, orange: logical qubits) while edges represent $CZ$ gates. Right side: Example of logical encoding using a depth-3 tree graph. (d) Quantum communication scheme using the RGS protocol. Matter qubits are also included.}
    \label{fig_both}
\end{figure*}

Several QR approaches have been proposed and can be divided into two main categories depending on the method used to propagate quantum information between two adjacent repeater nodes~\cite{Muralidharan2016}.
The first approach proposed, referred to here as a "memory-based" approach, relies on heralded-entanglement generation~\cite{Duan2001, Childress2006, Hartmann2007, Collins2007, Sangouard2011, Vinay2017, Rozpcedek2019, Bhaskar2020, Khatri2019} between two quantum memories (QM) situated at adjacent repeater nodes. Examples of memory-based schemes are depicted schematically in Figs.~\ref{fig_both}(a),(b).
The heralded entanglement generally succeeds when a detector situated at an intermediate node measures photons emitted by the QMs~\cite{Cabrillo1999, Barrett2005, Duan2004}.
A classical signal carrying the outcome of an entanglement generation attempt must be sent back to the QMs, thus limiting the repetition rate of these protocols.
When successful, entanglement swapping transfers the entanglement through these repeater nodes.

The second category of repeater~\cite{Jiang2009, Munro2012, Muralidharan2014, Azuma2015, Ewert2016, Hasegawa2019, Li2019} relies on logically-encoded multi-photon states~\cite{Varnava2006, Ewert2016}, resistant both to photonic losses and errors, to transfer quantum information across a network. One particularly promising approach of this type was put forward in Ref.~\cite{Azuma2015}, which proposed an all-optical QR protocol based on repeater graph states (RGS). An example of an RGS with logical encoding is shown in Fig.~\ref{fig_both}(c), while the corresponding communication protocol is summarized in Fig.~\ref{fig_both}(d) and will be detailed later. Ref.~\cite{Azuma2015} also proposed a method to construct these states probabilistically using single-photon sources, linear optics, and detectors~\cite{Browne2005}.
However, this generation procedure requires about $10^6$ single-photon sources per repeater node to just barely outperform direct fiber transmission~\cite{Pant2017}. These findings suggest that the RGS protocol may be well beyond the reach of current and near-future technological capabilities.

However, it remains unclear whether the RGS protocol can become more feasible if alternative methods to create the states are used. Ref.~\cite{Lindner2009} proposed a deterministic generation procedure to produce a linear graph state using one quantum emitter which emits spin-entangled photons~\cite{Saavedra2000}; this procedure was later demonstrated experimentally~\cite{Schwartz2016}. Ref.~\cite{Economou2010} showed theoretically that emitters that undergo entangling gates can also be used to generate two-dimensional cluster states. Refs.~\cite{Buterakos2017,Russo2019} built on this protocol and demonstrated that entanglement between emitters can be harnessed for the generation of more complex photonic graph states. These ideas have been extended further to general prescriptions and to protocols tailored to specific physical systems~\cite{Pichler2017, Russo2018, Chan2018, Gimeno2019}. Ref.~\cite{Buterakos2017} introduced a protocol for producing an arbitrary-sized RGS using only a few matter qubits, thus significantly decreasing the resource overhead required for RGS generation, making it deterministic in principle.
Using this generation technique, the RGS protocol only necessitates a few-qubit processor at each repeater node, which would ease its practical implementation compared to  other error-correction-based proposals that generally require several hundreds of qubits per repeater node~\cite{Munro2012, Muralidharan2014} (with the notable exception of Ref.~\cite{Borregaard2020} which uses techniques introduced in Ref.~\cite{Buterakos2017} and requires deterministic spin-photon Bell measurements). Although the RGS protocol with deterministic state generation seems promising, a systematic and detailed evaluation of its performance and resource requirements has not been carried out.

In this paper, we compare the resource-efficiency---characterized by the achievable secret key rate per matter qubit---of this protocol to direct fiber transmission and to QR schemes based on memories and heralded entanglement generation.
We first show that the rate per matter qubit has a fundamental upper bound in the case of memory-based QRs. We then review the RGS protocol and how RGSs can be generated using a few matter qubits. We evaluate the performance of this scheme, show that its rate per matter qubit does not have a theoretical upper-bound, and find the conditions under which it outperforms both the repeater-less and the memory-based QR approaches. These conditions depend sensitively on the speed with which two-qubit gates between the matter qubits can be executed and on the collection and detection efficiencies of the photons emitted by these matter qubits.

\section{Upper bound on rate for memory-based repeater schemes}
\label{sec_QM}

    In this section, we show that there is a theoretical upper bound, $R^{(QM)}_{\mathrm{max}}$, on the rate per matter qubit for protocols based on quantum memories and heralded entanglement generation. In such protocols, the total distance $L$ between Alice and Bob is divided into smaller distances $L_0$ by $N_{QR} = L/L_0 -1$ repeater nodes. Quantum memories at adjacent repeater nodes are entangled via a heralded entanglement procedure (see Fig.~\ref{fig_both}(a)). When a repeater node shares entanglement connections with its two adjacent nodes, entanglement swapping is performed on the two memory qubits within that node to create a direct entanglement connection between memories on the adjacent nodes. This procedure can be repeated until Alice and Bob share an entangled qubit pair.

    It is clear that creating an entangled pair between Alice and Bob requires generating an entanglement connection between each adjacent pair of repeaters. This means that the overall protocol rate $R^{(QM)}$ is limited by the entanglement generation rate $\langle T_{\mathrm{ent}} \rangle^{-1}$ between two adjacent repeaters. Here, we use this fact to derive an upper bound on the rate per matter qubit for QR protocols that use quantum memories and heralded entanglement generation.

    To determine an upper bound on $\langle T_{\mathrm{ent}} \rangle^{-1}$, we focus for concreteness on the protocol presented in Fig.~\ref{fig_both}(a), which uses two quantum memories per node.
    A quantum memory should emit a photon that is maximally entangled with one of its degrees of freedom.
    Two photons generated at adjacent repeater nodes arrive at the same measurement node situated halfway between the two repeaters, where they are measured in a Bell state basis.
    Because a photon Bell state measurement using only linear optics succeeds with probability at best $1/2$ (without ancillary qubits or QND measurements~\cite{Lloyd2001, Kim2001, Kim2002,Grice2011, Ewert2014,  Wein2016}), the overall success probability of the distant heralded entanglement generation is $P_{\mathrm{ent}} \leq 1/2$.
    It is worth mentioning that a method for achieving heralded entanglement generation with higher success probability has been proposed~\cite{Martin2019}, but its efficacy is restricted to qubits separated by a short distance, so we exclude this from our analysis (see Supplementary Materials for more details).
    A classical signal must then inform the repeater nodes of the success or failure of the Bell state measurement. Because the distance from the measurement node to the repeater nodes is $L_0/2$, this means that the overall heralded entanglement generation attempt takes total time $T_{\mathrm{trial}} \geq L_0/c$. This includes the time $L_0/(2c)$ for the single-photon transfer from the repeater to the measurement node and also the time $L_0/(2c)$ for the classical signaling in the opposite direction. Here, we are neglecting the time it takes to prepare and pump the quantum memories. Throughout this protocol, a QM can be maximally entangled with at most one other qubit (either a photon or another QM).
    Therefore, it cannot emit another spin-entangled photon before receiving the classical signal carrying the information about the success or failure of the Bell measurement, hence limiting the repetition rate of the protocol. To generate an entanglement connection between the two adjacent nodes, the procedure must be repeated on average $P_{\mathrm{ent}}^{-1}$ times. The entanglement generation rate is therefore $\langle T_{\mathrm{ent}} \rangle^{-1} = P_{\mathrm{ent}} / T_{\mathrm{trial}} < c/(2L_0)$.
    We derived this result for a specific heralded entanglement generation protocol but it also holds for all known protocols~\cite{Cabrillo1999, Duan2004, Barrett2005} (see Supplementary Materials).

    \begin{table}
        \begin{center}
            \begin{tabular}{ |p{1.6cm}p{7.0cm}|}
        \hline
        Notation & Definition\\
        \hline
        $L$ & Total distance between Alice and Bob.\\
        $L_0$ & Distance between adjacent nodes.\\
         $N_{QR}$ & Number of repeater nodes.\\
        $2m$ & Number of arms of the RGS.\\
        $\vec{b}$ & Branching vector of an error-correction tree ($\vec{b}=(b_0, b_1,\ldots,b_{n-1})$).\\
        $T_{CZ}$ & $CZ$ gate time.\\
        $\eta_t(l)$ & Transmission of a fiber of length $l$ (\(\eta_t(l) = \exp(-l/L_{\mathrm{att}})\)).\\
        $\eta_c, \eta_d$ & In-fiber collection and detection efficiencies of photons.\\
        $L_{\mathrm{att}}$, $c$ & Attenuation distance of the fiber and speed of light in a fiber. (\(L_{\mathrm{att}} \approx 20 \kilo\meter\))\\
        $t_{\mathrm{att}}$ & Average time of flight of photons in the fiber ($t_{\mathrm{att}} = L_{\mathrm{att}}/c$).\\
        $T_{RGS}$ & Generation time of an RGS.\\
        $R, R_m$ & Rate and rate per matter qubit of the protocol.\\
        $\epsilon$ & Single-photon error rate.\\
        \hline
        \end{tabular}
        \end{center}
        \caption{Table of notations}
        \label{table3}
    \end{table}

    From these results, we can show that the rate per matter qubit (where the number of matter qubits is $N_m = 2 (N_{QR} + 1) = 2L/L_0$) has a theoretical upper bound, $R^{(QM)}_{\mathrm{max}}$:
    \begin{equation}
        \frac{R^{(QM)}}{N_m} \leq \frac{\langle T_{\mathrm{ent}} \rangle^{-1}}{N_m} \leq \frac{c}{4L} =R_{\mathrm{max}}^{(QM)}.
    \end{equation}

    This theoretical upper bound also holds if there are more than two QMs at each repeater node (see Fig.~\ref{fig_both}(b)) as the rate would linearly increase with the number of matter qubits. Therefore, we have derived a general theoretical upper bound for memory-based protocols based on heralded entanglement generation. It is worth noting that the fundamental reason for this upper bound comes from the need for classical signaling in these protocols. Such classical signaling is not required for RGS protocols, enabling them to surpass this limit, as we show below. In the Supplementary Materials, we also show that a tighter bound, $R_{\rm max}^{(QM)}=c/7L$, can be obtained for memory-based schemes in which there are two quantum memories per repeater node, and heralded entanglement swapping is used.

    We emphasize that the upper bound derived in this section holds for QR protocols that are based on quantum memories and distant heralded entanglement generation. This corresponds to the first and second generations of QRs, as categorized in Ref.~\cite{Muralidharan2016}. Consequently, in the following, the RGS-based protocol will be compared only to these categories of QRs, for which the performance is limited by classical signaling.

\section{Rate of the RGS protocol with deterministic graph state generation}

In this section, we review the RGS protocol as introduced in Ref.~\cite{Azuma2015} and the deterministic generation of RGSs using a few matter qubits as proposed in Ref.~\cite{Buterakos2017}. We show how the rate of the RGS protocol depends on various parameters in the case where deterministic state generation methods are used.

\subsection{RGS protocol and rate}

    An RGS is a quantum state $\ket{G}$ that can conveniently be represented in the form of a graph $G = (V, E)$ with $V$ vertices and $E$ edges. Each vertex corresponds to a photonic qubit prepared in the $\ket{+}$ state, and each edge corresponds to the application of a $CZ$ gate between the two qubits it connects:
    \begin{equation}
        \ket{G} = \prod_{(i,j) \in E} CZ_{ij} \ket{+}^{\otimes V}.
    \end{equation}
    An example of the graph representing an RGS is shown in Fig.~\ref{fig_both}(c). These states include $2m$ inner photonic qubits that are referred to as the first-leaf qubits. All the first-leaf qubits are fully connected to each other and each of them is also connected to one additional qubit, referred to as a second-leaf qubit. The first-leaf qubits are logically-encoded using tree graph states; further details on this are given below.

    In an RGS protocol (see Fig.~\ref{fig_both}(d)), the distance $L$ separating Alice and Bob is also divided into smaller steps $L_0$ by $N_{QR} = L/L_0 - 1$ source nodes where the RGSs are created. The RGS is divided into two equal parts, each containing $m$ arms, and one part is sent to the left adjacent measurement node and the other to the right. Thus, half of one RGS meets half of another RGS at each measurement node, where each second-leaf qubit from one of the half-RGSs undergoes a Bell measurement with its counterpart from the other half-RGS.
    Further details about this entanglement swapping procedure at the measurement node are given later, but it is important to note that the RGS protocol does not use quantum memories at all and thus cannot store the information. This implies that an entanglement connection between Alice and Bob should be realized in only one trial with a probability $P_{AB}$ much higher than direct fiber transmission: $P_{AB} \gg \eta_t(L)$.
    The generation of an entanglement connection between Alice and Bob requires the realization of successful entanglement connections between all the adjacent RGSs. It is important to note that the measurements performed at each measurement node do not require information from other measurement nodes so that, in contrast to memory-based approaches, the RGS protocol does not require any classical signaling while entanglement is being extended through the network. Classical signaling is needed only once at the end of the protocol to recover the Pauli frame of the Bell pair shared by Alice and Bob, i.e. to determine which local Pauli rotations Alice and Bob's qubits should undergo.
    This means that in the RGS protocol, it is not necessary to wait for any classical signaling before proceeding to generate the next batch of RGSs needed to create the next Bell pair shared between Alice and Bob (see Supplementary Materials for more details). Consequently, unlike the memory-based scheme, the rate of the RGS protocol does not depend on the time it takes for the photon to get from one node to the next. It is limited only  by the generation time $T_{RGS}$ of an RGS:
    \begin{equation}
        R^{(RGS)} = \frac{{P_{RGS \leftrightarrow RGS}}^{L/L_0}}{T_{RGS}},
        \label{eq_rate}
    \end{equation}
    with $P_{RGS \leftrightarrow RGS}$ the probability to generate an entanglement connection between two RGSs. The main notations used in this work are defined in Table~\ref{table3}.

    \begin{figure*}[!ht]
        \centering
        \includegraphics[width=16cm]{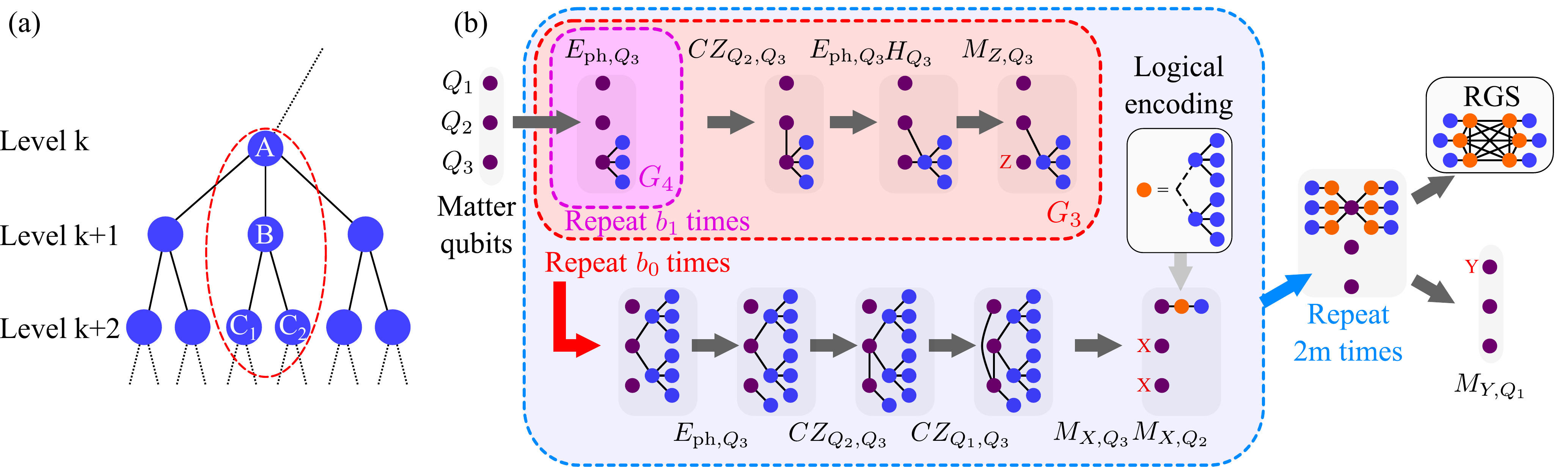}
        \caption{ (a) Indirect measurement of qubit A at the level $k$ using a stabilizer based on qubit B at level $k+1$. (b) Protocol for the deterministic generation of an RGS with logical encoding using matter qubits.}
        \label{fig_generation}
    \end{figure*}

    A successful entanglement link can be generated if at least one of the Bell measurements at a measurement node succeeds. In that case, the two first-leaf qubits attached to the second-leaf qubits that underwent the successful Bell measurement are measured in the $X$ basis, while the remaining $2m-2$ first-leaf qubits are measured in the $Z$ basis. The $X$ measurements transfer the entanglement connection to the next two adjacent measurement nodes, while the $Z$ measurements disentangle all the excess qubits associated with failed Bell measurements. All these first-leaf qubit measurements must be successful in order to reliably create an entanglement link. Therefore, the probability to successfully create an entanglement link between two RGSs is given by:
    \begin{equation}
        \begin{aligned}
            P_{RGS \leftrightarrow RGS} & = \left(1 - \left(1 - P_{\mathrm{Bell}}\right)^m\right) \\
            & \times \mathrm{Pr}(M_{X,\ell})^2 \mathrm{Pr}(M_{Z,\ell})^{2m-2},\label{eq:PRGStoRGS}
        \end{aligned}
    \end{equation}
    where $P_{\mathrm{Bell}} = {P_{\mathrm{ph}}}^2 / 2$ is the probability of a successful Bell measurement. This depends on $P_{\mathrm{ph}} = \eta_c \eta_d \eta_t(L_0/2)$, which is the probability that a single photon is emitted and collected into the fiber ($\eta_c$), is transmitted to the measurement node ($\eta_t(L_0/2)$) and detected ($\eta_d$). $\mathrm{Pr}(M_{X,\ell})$ and $\mathrm{Pr}(M_{Z,\ell})$ are the probabilities that the logical $X$ and $Z$ measurements on the first-leaf qubits succeed.
    Note that if the first-leaf qubits were not logically encoded, we would have $\mathrm{Pr}(M_X)^2 \mathrm{Pr}(M_Z)^{2m-2} = {P_{\mathrm{ph}}}^{2m} \leq \eta_t(L_0)$, and it would be impossible to have an advantage over direct fiber transmission. Therefore, the loss-tolerance of logically-encoded qubits is crucial for this protocol. Next, we review this encoding, which was introduced in Refs.~\cite{Varnava2006} and~\cite{Azuma2015}.

    \subsection{Loss-tolerance with tree graph states}

    We review how the probabilities $\mathrm{Pr}(M_{X,\ell})$ and $\mathrm{Pr}(M_{Z,\ell})$ depend on the single-photon transfer probability $P_{\mathrm{ph}}$ and on the shape of the tree graph state used for the logical encoding.
    Ref.~\cite{Varnava2006} demonstrated that this encoding remains loss-tolerant as long as $P_{\mathrm{ph}}$ is above $50\%$.

    We consider the calculation of the probabilities of successful measurements of the logically encoded qubits in the presence of loss errors on a tree graph state. A tree is characterized by its branching vector $\vec{b} = (b_0, b_1,..., b_{n-1})$ (see Fig.~\ref{fig_both}(c)), which describes the connectivity between the different levels of the tree.
    To perform a $Z$ measurement $M_{Z,k}$ on a qubit at level $k$, it is possible to either perform a direct measurement on this qubit (with success probability $P_{\mathrm{ph}}$) or, if it fails (with probability $1 - P_{\mathrm{ph}}$), perform an indirect measurement (with probability $r_k$). Thus, the overall success probability of a $Z$ measurement at level $k$ is:
    \begin{equation}
        \mathrm{Pr}(M_{Z,k}) = P_{\mathrm{ph}} + (1 - P_{\mathrm{ph}}) r_k.
    \end{equation}

    To perform an indirect measurement on a qubit (call it A) at level $k$, one can use the stabilizing property of a graph state~\cite{Hein2004}. It is possible to deduce the outcome of the $Z$ measurement on A by performing an $X$ measurement on another qubit (B) at level $k+1$ and a $Z$ measurement on all the qubits, $C_i$, that are in the neighborhood of B at level $k+2$ (see Fig. \ref{fig_generation}(a)). This works because of the invariance of graph states when they are acted upon by their stabilizers:
    \begin{equation}
        \begin{aligned}
            \ket{G}&  = X_B \tens{j \in N(B)} Z_j \ket{G} \\
            & = X_B \otimes Z_A \tens{i\in \{1, b_{k+1}\}} Z_{C_i} \ket{G},
        \end{aligned}
    \end{equation}
    so we have:
    \begin{equation}
        Z_A \ket{G} =X_B \tens{i\in \{1, b_{k+1}\}} Z_{C_i} \ket{G}.
    \end{equation}

    A single indirect measurement has a success probability $s_k$. Note, however, that the tree structure allows $b_k$ indirect measurement attempts, and only one needs to  succeed to indirectly measure a qubit at level $k$. So the probability that at least one indirect measurement succeeds is
    \begin{equation}
        r_k = 1 - (1 - s_k)^{b_k},
    \end{equation}
    with
    \begin{equation}
        s_k = P_{\mathrm{ph}} \mathrm{Pr}(M_{Z,k+2})^{b_{k+1}}.
    \end{equation}
It is thus possible to derive the success probability of a measurement recursively, given that the qubits at the lowest level can only be measured directly: $\mathrm{Pr}(M_{Z,n}) = P_{\mathrm{ph}}$. Logical measurements in the $X$ or $Z$ basis are  given by~\cite{Azuma2015}:
    \begin{equation}
    \begin{aligned}
        \mathrm{Pr}(M_{X,\ell}) & = r_0, \\
        \mathrm{Pr}(M_{Z,\ell}) & = \left(P_{\mathrm{ph}} + (1 - P_{\mathrm{ph}}) r_1 \right)^{b_0} \\
        & = \mathrm{Pr}(M_{Z,1})^{b_0}.
    \end{aligned}
    \end{equation}
    It is interesting to note that logical encoding with tree graph states can also correct single-qubit errors, as shown in Ref.~\cite{Azuma2015} and described in the Supplementary Materials. This will be used later when we evaluate the sensitivity of the RGS protocol to errors.

    \subsection{Generation of an RGS}
    The achievable rate between Alice and Bob also depends on the repetition rate of the protocol, which is given by the generation time of the RGS. Because it is impossible to realize deterministic two-qubit gates on photons with linear optics, such a graph state can either be generated probabilistically by the recursive fusion of smaller graphs using linear optics and Bell state measurements as shown in Ref.~\cite{Azuma2015} and~\cite{Pant2017}, or deterministically using a few matter qubits as shown in Ref.~\cite{Buterakos2017}. We now review the latter.

    An arbitrary-sized RGS can be generated deterministically by following a given sequence based on four operations on matter qubits: the emission of a photon maximally entangled with the matter qubit $E_{\mathrm{ph}}$, the Hadamard gate $H$, measurements in the Pauli bases $M_X$, $M_Y$, $M_Z$ and the $CZ$ gate.
    The generation of an RGS with $2m$ arms and a tree graph encoding with branching vector $\vec{b}=(b_0, b_1, ..., b_{n-1})$ requires $n + 1$ matter qubits $Q_1$, ..., $Q_{n+1}$,  and is given by the sequence (see also Fig.~\ref{fig_generation}(b))

\begin{equation}
    \begin{aligned}
    & M_{Y,{Q_1}}(M_{X,Q_3} M_{X,Q_2} CZ_{Q_1,Q_3} CZ_{Q_2,Q_3} E_{\mathrm{ph},Q_3} {G_3}^{b_0})^{2m} \\
& \mathrm{with} \; G_{k} = M_{Z,Q_{k}}  H_{Q_{k}} E_{\mathrm{ph},Q_{k}}  CZ_{Q_{k-1}, Q_{k}} {G_{k+1}}^{b_{k-2}} \\
& \mathrm{and} \; G_{n+ 2} = E_{\mathrm{ph},Q_{n+ 1}} ,
    \end{aligned}
\end{equation}
 where, for simplicity, we have omitted the single photonic qubit rotations.

The overall generation time of an RGS using this procedure is therefore
    \begin{equation}
        \begin{aligned}
             T_{RGS} & = 2m \left(1 + f(\vec{b}, n-1)\right) T_{E_{\mathrm{ph}}} +T_M\\
             &  + 2m \left(2 + f(\vec{b}, n-2)\right) (T_M + T_{CZ}) \\
             & + 2m   f(\vec{b}, n-2) T_{H},  \\
        \end{aligned}
    \end{equation}
    with $f(\vec{b}, k) = \sum_{i=0}^{k} \prod_{j=0}^{i} b_j$ and with $T_{E_{\mathrm{ph}}}$, $T_M$, $T_H$ and $T_{CZ}$  the times for photon emission, matter qubit measurement, Hadamard and $CZ$ gates, respectively.

    In the following, we make the realistic assumption that the $CZ$ gate time $T_{CZ}$ is much longer than the durations of the other operations, and so we set $T_{H} = T_{M} = T_{E_{\mathrm{ph}}} = 0$ for simplicity. With this assumption, the generation time $T_{RGS}$ only depends on the number of $CZ$ gates and their duration:
    \begin{equation}
        T_{RGS} = 2m \left( 2 + \sum_{k=0}^{n-2}{\prod_{j=0}^{k} b_j} \right) T_{CZ}.
        \label{eq_t_rgs}
    \end{equation}
    This is for an RGS with $2m$ arms and a logical tree encoding with branching vector $\vec{b} = (b_0, ..., b_{n-1})$. In the following, we will assume a depth-two tree graph state ($\vec{b} = (b_0, b_1)$), so that the full RGS can be generated with only three matter qubits ($n=2$). The total number of matter qubits in the network is therefore $N_m=(L/L_0-1)(n+1) = 3(L/L_0-1)$.

    \section{Performance comparisons}

    We now evaluate the performance of the RGS protocol when it is generated deterministically by a few matter qubits. We compare our results to the direct fiber transmission limit derived by Refs.~\cite{Takeoka2014, Pirandola2017} and to the memory-based upper bound $R_{\mathrm{max}}^{(QM)} = c/4L$ found in Sec.~\ref{sec_QM}.

    \subsection{Optimizing the RGS protocol}

    \begin{figure}[!ht]
        \centering
        \includegraphics[width=8cm]{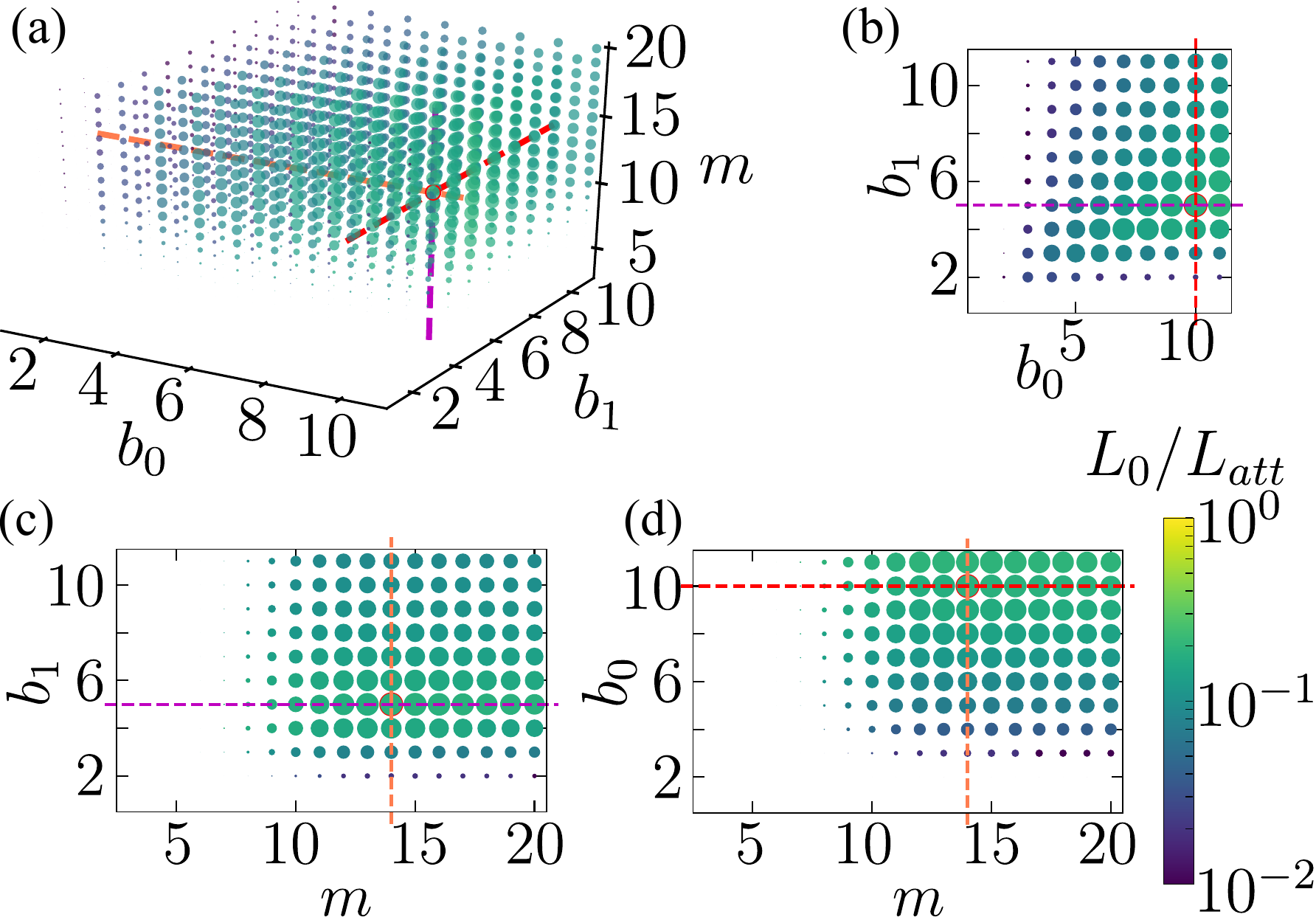}
        \caption{Maximum achievable rate per matter qubit $R_m^{(RGS)}T_{CZ}$ (here normalized by the parameter ${T_{CZ}}^{-1}$) and optimal node separation $L_0$ for the RGS protocol with depth-2 logical tree encoding with the total distance fixed at $L=50L_{\mathrm{att}}$ for a range of RGS parameters $b_0$, $b_1$, $m$. (a) Three-dimensional plot showing optimal rate and node separation as a function of $b_0$, $b_1$ and $m$. For each point in the plot, the value of $L_0$ is optimized to maximize $R_m^{(RGS)} T_{CZ}$. Point sizes represent maximized $R_m^{(RGS)} T_{CZ}$ values while point colors represent the optimal values of $L_0$ (indicated with the color scale). The RGS parameters that achieve the largest value of $R_m^{(RGS)} T_{CZ}$ in this case are indicated with dashed lines. The corresponding maximal $R_{\rm max}^{(RGS)}T_{CZ}$ is given in Table~\ref{table}. (b,c,d) Three different orthogonal two-dimensional slices of the plot shown in panel (a).}
        \label{fig_optimization_tree_RGS}
    \end{figure}

          \begin{table}
              \begin{center}
                  \begin{tabular}{ |p{1.2cm}|p{1.2cm}|p{0.6cm}|p{0.6cm}|p{0.6cm}|p{1.8cm}| }
              \hline
              $L/L_{\mathrm{att}}$ & $L_0/L_{\mathrm{att}}$ & $m$ & $b_0$ & $b_1$ & $R_{\rm max}^{(RGS)} T_{CZ}$\\
              \hline
              \hline
              10 & 0.23 & 11 & 8 & 4 & $2.4\times 10^{-5}$ \\
              \hline
              25 & 0.21 & 13 & 10 & 5 & $6.8\times 10^{-6}$\\
              \hline
              50 & 0.19 & 14 & 10 & 5 & $2.8\times 10^{-6}$\\
              \hline
              100 & 0.17 & 15 & 10 & 5 & $1.1\times 10^{-6}$\\
              \hline
              150 & 0.15 & 16 & 10 & 5 & $6.8\times 10^{-7}$\\
              \hline
              \end{tabular}
              \end{center}
              \caption{Optimal RGS parameters $m$, $b_0$, $b_1$ and node separation $L_0$ for several different total network distances $L$. Here, $L_{\rm att}$ is the attenuation length of optical fibers.}
              \label{table}
          \end{table}

    In the following, we show how the RGS protocol parameters can be optimized to maximize the overall rate per matter qubit, $R_m^{(RGS)}$, or in the presence of errors, the secret key rate per matter qubit.
    For the moment, we assume perfect photon collection and detection efficiencies ($\eta_c \eta_d = 1$) and no single-photon errors ($\epsilon=0$); we take these effects into account later on. From Eqs.~\eqref{eq_rate}, \eqref{eq:PRGStoRGS}, and~\eqref{eq_t_rgs}, the achievable rate per matter qubit $R_m^{(RGS)}$ of the RGS protocol for a total distance $L$ is inversely proportional to the $CZ$ gate time $T_{CZ}$, but it depends non-linearly on the separation distance, $L_0$, between two RGS nodes and the RGS shape (number of arms $2m$ and the tree branching vector $\vec{b} = (b_0, b_1)$). Therefore, for each choice of the total distance $L$, there is a certain node separation and RGS shape that maximize the achievable rate.

    The optimization of the rate per matter qubit for a total distance $L = 50L_{\mathrm{att}}$ ($\approx 1000\; \kilo\meter$) is shown in Fig.~\ref{fig_optimization_tree_RGS}. The position of each point corresponds to a specific RGS shape, the color of the point indicates the node separation $L_0$ that optimizes the rate for that shape, and the size of the point represents the maximal rate per matter qubit for these parameters.
    This optimization converges, allowing us to extract the optimal RGS shape and distance $L_0$ for this particular choice of the total distance $L$. The optimization can be repeated for various choices of $L$, and the extracted optimal parameters are recorded in Table~\ref{table}.

\begin{figure*}
        \centering
        \includegraphics[width=18cm]{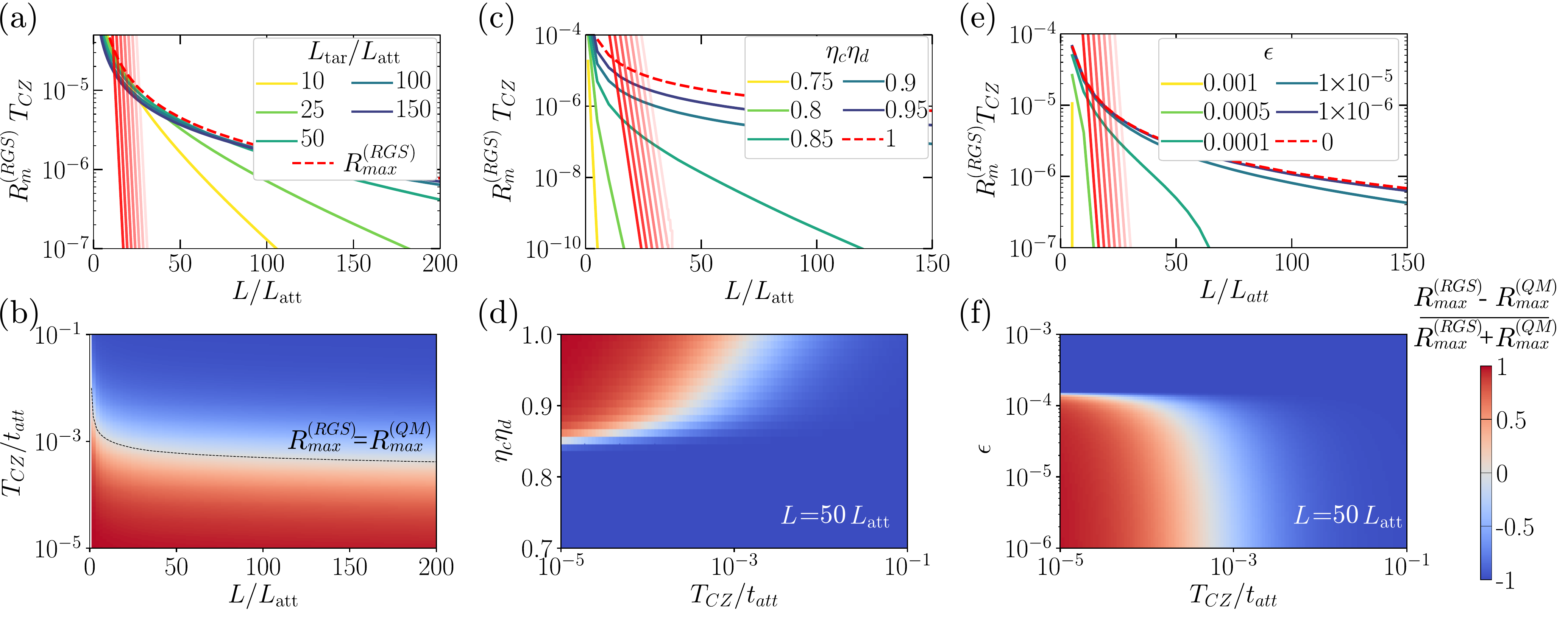}
        \caption{
        (a) The maximum achievable secret key rate per matter qubit as a function of the total distance $L$ for the RGS protocol with depth-2 logical tree encoding compared to direct fiber transmission. The dashed red line is the rate for the RGS protocol, with all RGS parameters and $L_0$ optimized for each value of $L$. The seven red curves correspond to rates for direct fiber transmission \cite{Pirandola2017} for seven different choices of the single-photon source repetition rate: $10^i {T_{CZ}}^{-1}$ (for i=0,...,6).  The solid yellow, green, and blue curves show the performance of the RGS protocol when $L$ is changed while keeping all RGS parameters and $L_0$ fixed to the optimal values obtained for total distance $L_{\rm tar}$.
        (b) Relative difference between the maximum achievable rates of the RGS protocol, $R^{(RGS)}_{\mathrm{max}}$, and memory-based protocols, $R^{(QM)}_{\mathrm{max}}$, as a function of gate time $T_{CZ}$ and total distance $L$. The red region corresponds to $R^{(RGS)}_{\mathrm{max}}>R^{(QM)}_{\mathrm{max}}$.
        (c,d) Optimized rate of the RGS protocol for different values of photon loss probability $\eta_c\eta_d$ compared to (c) direct fiber transmission and (d) memory-based limit. (e,f) Similar to (c,d) but now showing results for the single-photon error probability $\epsilon$. In (c,e), the RGS rate is maximized for each value of $L$.}
        \label{fig_optimization_L}
    \end{figure*}

We compare the RGS protocol to direct fiber transmission in Fig.~\ref{fig_optimization_L}(a). The maximum achievable rate per matter qubit for the RGS protocol, $R^{(RGS)}_{\mathrm{max}}$, is shown as a function of the total distance $L/L_{\rm att}$.
In the case of direct transmission, we show seven different curves corresponding to the achievable rate for seven different values of the single-photon source repetition rate. We see that the RGS protocol outperforms direct transmission with the highest repetition rate for $L\gtrsim30L_{\rm att}$. In the same figure, we also show how well the protocol works if we keep the optimized parameters fixed and change the total distance $L$. To demonstrate this, we fix the RGS parameters and node separation $L_0$ to the values that optimize the rate for a total distance $L_{\mathrm{tar}}$. We then adjust $L$ away from $L_{\mathrm{tar}}$ without changing the RGS parameters or $L_0$, and we calculate the new rate for each value of $L$. In Fig.~\ref{fig_optimization_L}(a), we show the resulting rates as a function of $L/L_0$ for five different choices of $L_{\mathrm{tar}}$. We see that the RGS protocol continues to work well over a broad range of total distance $L$ when we use parameters that are optimized for a large total distance $L_{\mathrm{tar}}$.

    For long distances $L$ in the range $50L_{\mathrm{att}} - 200L_{\mathrm{att}}$, $R^{(RGS)}_{\mathrm{max}}$ scales approximately as $L^{-1.27}$, and thus the scaling with distance is slightly worse than the upper bound scaling for memory-based schemes, $R^{(QM)}_{\mathrm{max}}\sim L^{-1}$, obtained in Sec.~\ref{sec_QM}.
    However, contrary to the memory-based schemes, there are no fundamental upper limits imposed by classical signaling on the maximal achievable rate $R^{(RGS)}_{\mathrm{max}}$, as the latter is inversely proportional to the $CZ$ gate time, $T_{CZ}$. If $T_{CZ}$ is sufficiently small, then the RGS protocol should surpass the upper bound of memory-based schemes. This is illustrated in  Fig.~\ref{fig_optimization_L}(b), where the red region of the plot indicates the regime where the RGS protocol outperforms memory-based schemes.

    From our calculations, it seems that the maximum achievable rate per matter qubit, $R^{(RGS)}_{max}$, is in principle unbounded since it is inversely proportional to the $CZ$ gate time. We should recall however that these results are based on the assumption that the $CZ$ gate takes much longer than the other operations $O$ made on the matter qubits: $T_{CZ} \gg T_{O}$ . Decreasing $T_{CZ}$ initially increases the rate, but if $T_{CZ}$ becomes small enough, neglecting the durations of other operations eventually becomes invalid. Consequently, to increase the rate further, not only the $CZ$ gate time but all the operation times should be reduced simultaneously. If the durations of all these operations could be made arbitrarily small, then the rate per matter qubit would increase to infinity. In practice however, the operation times could also have an intrinsic lower bound that limits the performance of the RGS protocol, but these lower bounds would depend on the specific system on which the protocol is implemented, while the limit imposed by classical signaling is much more stringent and general.

    For distances below $200L_{\mathrm{att}} \approx 4000\;\kilo\meter$, a gate time $T_{CZ}$ below $6\times10^{-4} t_{\mathrm{att}} \approx 60\;\nano\second$ is sufficiently short to outperform any memory-based scheme.
    For this range of total distance, the optimal value of the node separation $L_0$ ranges from $0.15-0.19 L_{\mathrm{att}} \approx 3-3.8\;\kilo\meter$.
    As an illustration of the RGS performance, for $L = 1000 {\rm \;  km}$ and $T_{CZ} = 10 {\rm \;  ns}$, the total rate is $R = 220 {\rm \; kHz} $ for $N_m = 786$ matter qubits used,
   leading to $R_{\rm max}^{(RGS)}=276 {\rm \;  Hz}$ per matter qubit, while $R_{\rm max}^{(QM)} = 50 {\rm \; Hz}$ per matter qubit for this distance.

    \subsection{Sensitivity to errors}

    So far, we have considered the generation of a perfect RGS, which is a pure entangled state of many photons. Generating such a perfect state is not feasible experimentally, and so we need to evaluate the sensitivity of the RGS protocol to errors.
    To do so, we now consider single-photon loss and single-photon errors in our optimization process.

    The single-photon loss (other than the fiber losses) depends on the probability that an emitted photon is neither collected into the fiber nor detected; this corresponds to the case where $\eta_c \eta_d \neq 1$. Fig.~\ref{fig_optimization_L}(c) shows the optimized rate per matter qubit for different values of $\eta_c \eta_d$.
    Similarly, Fig.~\ref{fig_optimization_L}(d) compares the performance of the RGS protocol with the upper bound for memory-based protocols, $R_{\mathrm{max}}^{(QM)}$, as a function of $T_{CZ}$
    and $\eta_c \eta_d$ for a total distance $L = 50 L_{\mathrm{att}}$.
    These results show that the photon losses must be below $15\%$ ($\eta_c \eta_d > 85\%$) in order for the RGS protocol to outperform both the direct fiber transmission and memory-based protocols. This rather stringent  requirement is mainly due to the loss-tolerance of the tree graph state, which only works when the total photon loss (including fiber losses) is below $50 \%$.

    Apart from the photon losses, the RGS protocol also depends on other kinds of errors that reduce the fidelity $F_{AB}$ of the final entangled photon pair shared by Alice and Bob. These errors include, but are not limited to, single-photon measurement errors, photon Bell-measurement errors, and depolarization errors. In the present case where the RGS is generated using matter qubits, the limited coherence time of the matter qubits should also limit the fidelity of the entangled photons.
    For QKD applications, the final secret key rate depends on the total rate $R$ and the fidelity $F_{AB}$~\cite{Shor2000, Scarani2009}:
    \begin{equation}
        R_{skr}=R (1 - 2h(F_{AB})),
        \label{eq_skr}
    \end{equation}
    where $h$ is the binary entropy function:
    \begin{equation}
        h(F) =  - F \log_2(F) - (1 - F) \log_2(1-F).
    \end{equation}
    In the remainder of this paper, we now use $R_m^{(RGS)}$ to refer to the secret key rate per matter qubit, which coincides with the rate per matter qubit in the absence of errors. We model these imperfections by using a single-qubit error probability $\epsilon$ that affects all the photonic qubits. This model works well for qubit measurement errors as well as depolarization errors. In Ref.~\cite{Lindner2009}, it was also shown that decoherence errors on quantum emitters induce only local errors in the emitted cluster state.
    The detailed derivation of $F_{AB}$ is given in the Supplementary Materials. Fig.~\ref{fig_optimization_L}(e) compares the optimized secret key rate of the RGS protocol against that of direct fiber transmission as a function of the single-qubit error probability and the total distance $L$. Fig.~\ref{fig_optimization_L}(f) compares the performance of the RGS protocol to the upper bound for memory-based protocols for $L = 50 L_{\mathrm{att}}$. We find that the single-qubit error should be below $\epsilon < 10^{-4}$ to outperform both the direct fiber transmission and the memory-based protocol upper bound, $R^{(QM)}_{\rm max}$.

    One might expect a better tolerance to single-qubit errors since they are also corrected by the logical encoding of the 1\st-leaf qubits. This is not the case because the 2\nd-leaf qubits are not logically encoded and thus are very sensitive to errors. The dependence on single-qubit errors is actually similar to the protocols using heralded entanglement generation. It is also possible to use entanglement purification~\cite{Pan2001, Zwerger2012, Zwerger2016} to improve the fidelity $F_{AB}$ and maximize the secret key rate.

    From the maximum single-qubit error rate, we can derive an order-of-magnitude estimate for the coherence time requirements for the matter qubits.
    To do so, we use two strong assumptions. We first assume that the fidelity $F_{RGS}$ of an RGS depends on the ratio between the generation time $T_{RGS}$ of the RGS and the coherence time $T_2$ of the matter qubits:
    \begin{equation}
        F_{RGS} \approx e^{- \frac{3 T_{RGS}}{T_2}}
    \end{equation}
    The coefficient 3 corresponds to the number of matter qubits that degrade the RGS fidelity.
    Here, we have made the simplifying assumption that the main source of infidelity is the matter qubit decoherence, and we have assumed that the effect of this decoherence is to reduce the overall fidelity of the graph state by $e^{- T_{RGS}/T_2}$, which decays exponentially with the time during which the qubit is attached to the RGS ($T_{RGS}$) divided by its coherence time. This assumption may or may not be justified depending on the system used to implement the protocol, but we use it simply to get an order of magnitude estimate of the typical coherence time that is required for this protocol.
    We also assume that the errors are homogeneously distributed over the $N_{\mathrm{ph}}$ photons of an RGS, such that $F_{RGS} = (1-\epsilon)^{N_{\mathrm{ph}}}$.
    This assumption is consistent with Ref.~\cite{Lindner2009}, which shows that errors generated by a quantum emitter and imprinted onto the generated graph state are localized.
    With these assumptions, the coherence time should satisfy $T_2 \geq 2500 T_{CZ}$.

\section{Discussion}

\begin{table}
    \begin{center}
        \begin{tabular}{ |p{1.8cm}|p{1.2cm}|p{1.0cm}|p{1.2cm}|p{1.1cm}| }
    \hline
    \textbf{Matter qubit} & $\bm{T_{CZ}}$ & {\bm{$F$}} & $\bm{T_2}$ & $\bm{\beta}$\\
    \hline
    \hline
    Coupled QDs & \textcolor{blue}{1ns}~\cite{Greilich2011} \;~\cite{Kim2011, Gimeno2019} & \textcolor{red}{80\%} \;~\cite{Kim2011} & \textcolor{blue}{$3\micro\second$} \;~\cite{Greilich2007, Wang2012} & \textcolor{blue}{99.7\%} \; ~\cite{Najer2019}\\
    \hline
    defects (diamond) & \textcolor{orange}{$120 \nano\second$} \;~\cite{Fuchs2011,Solenov2013,Goldman2020}  & \textcolor{blue}{99.2\%} \; ~\cite{Rong2015} & \textcolor{blue}{0.6s} \; ~\cite{Bar-Gill2013} & \textcolor{blue}{99\%} \; ~\cite{Bhaskar2020} \\
    \hline
    Trapped ions & \textcolor{red}{$10\micro\second$} \;~\cite{Ballance2016} & \textcolor{blue}{99.9\%} \; ~\cite{Ballance2016} & \textcolor{blue}{10min} \;~\cite{Wang2017b} & \textcolor{blue}{89\%} \; ~\cite{Daiss2019}\\
    \hline
    \end{tabular}
    \end{center}
    \caption{State-of-the-art characteristics of candidate quantum emitters. $T_{CZ}$ is the duration of a $CZ$ gate, $T_2$ is the coherence time, $\beta$ is the single-mode photon emission probability, and $F$ the two-qubit gate fidelity.
    Blue (red) color means that the condition is (not) fulfilled with this system, while orange means that it is theoretically possible but has not been realized experimentally. Note that all these results were not achieved on the same system at the same time.}
    \label{table2}
\end{table}

In the previous section, we derived a set of requirements that need to be fulfilled in order for the RGS protocol to have a clear advantage over both memory-based QRs and repeater-less protocols:
\begin{enumerate}[(i)]
    \item $T_{CZ}\leq T_{lim}(L) \approx 60 \nano\second$,
    \item $T_2 \geq 2500 T_{CZ}$,
    \item $\eta_c \eta_d > 85 \%$.
\end{enumerate}
It is important to note that these bounds were obtained using RGSs with a depth-2 logical tree encoding. We may expect an increase in the rate per matter qubit if deeper trees were used but the improvement would certainly not be large as $T_{RGS}$ grows quickly with tree depth. Checking this directly is challenging because the process of optimizing the RGS protocol becomes significantly more complex. We leave this intensive numerical analysis to future work.

Table~\ref{table2} shows the current state-of-the-art for three candidate systems that can be used to generate RGSs. So far, the characteristics of the current systems are not good enough yet to exceed the memory-based resource-efficient upper bound $R^{(QM)}_{\mathrm{max}}$.
The main limitation comes from long two-qubit gate times, since fast and high-fidelity gates are critical for the performance of the RGS protocol. Currently, most of the two-qubit gates that have been experimentally demonstrated are slow because they rely on local interactions, such as the hyperfine coupling, which are generally weak. Efficient light-matter interfaces using cavities can also be used to realize deterministic, photon-mediated $CZ$ gates between two matter qubits~\cite{Solenov2013}, using spin-photon gates~\cite{Hu2008, Koshino2010, Rosenblum2011}.
The crucial advantage of this technique is that it only requires optical interactions, which are intrinsically much faster than hyperfine interactions.
Progress in the experimental realization of strong spin-dependent phase shifts has been made recently~\cite{Sun2016, Bhaskar2020, Wells2019, Androvitsaneas2019}.
The second condition (ii) will also be easier to achieve if the $CZ$ gate time is reduced.

The third condition (iii) requires the efficient collection of single photons emitted by a single quantum emitter. The collection of single photons needs to be facilitated by increasing the emission of photons into a single optical mode. This is generally realized using waveguides or cavities with single-mode emission probabilities $\beta$ that can reach $99.7\%$~\cite{Najer2019}. This single mode should then  be efficiently coupled to a single-mode fiber.
Finally, superconducting photon detectors can be used to achieve high detection efficiencies~\cite{Hadfield2009}.

In conclusion, we have identified the main requirements that must be met in order for all-photonic repeater protocols based on deterministically generated graph states to outperform memory-based protocols. We did this by finding the graph state structures that maximize the secret key rate and by comparing this to an upper bound on the rate for memory-based schemes that we derived. We found that two-qubit gate times and the efficient collection and detection of emitted photons are the most important factors in determining whether the all-photonic approach is superior to memory-based ones.
An interesting future direction would be to improve the protocol for the emitter-based generation of photonic graph states by incorporating atom-photon $CZ$ gates \cite{Rosenblum2011, Pichler2017, Zhan2020}, which would both simplify the procedure and decrease gate times, and thus enhance the performance of deterministic, all-photonic repeaters.

\section*{Acknowledgements}
We thank Yuan Zhan and Shuo Sun for their contribution in solving issues in the code.
This research was supported by the NSF (Grant No. 1741656) and by the EU Horizon 2020 programme (GA 862035 QLUSTER).

\section*{Methods}
The numerical model used for the results in this article is available here:

\url{https://github.com/Paulhilaire/performance_repeater_graph_state}.

\bibliographystyle{unsrtnat}
\bibliography{Biblio}

\begin{thebibliography}{85}
\providecommand{\natexlab}[1]{#1}
\providecommand{\url}[1]{\texttt{#1}}
\expandafter\ifx\csname urlstyle\endcsname\relax
  \providecommand{\doi}[1]{doi: #1}\else
  \providecommand{\doi}{doi: \begingroup \urlstyle{rm}\Url}\fi

\bibitem[Kimble(2008)]{Kimble2008}
H.~J. Kimble.
\newblock The quantum internet.
\newblock \emph{Nature}, 453\penalty0 (7198):\penalty0 1023--1030, 2008.
\newblock \doi{10.1038/nature07127}.

\bibitem[Wehner et~al.(2018)Wehner, Elkouss, and Hanson]{Wehner2018}
Stephanie Wehner, David Elkouss, and Ronald Hanson.
\newblock Quantum internet: A vision for the road ahead.
\newblock \emph{Science}, 362\penalty0 (6412):\penalty0 eaam9288, 2018.
\newblock \doi{10.1126/science.aam9288}.

\bibitem[Bennett and Brassard(1984)]{Bennett1984}
Charles~H Bennett and Gilles Brassard.
\newblock An update on quantum cryptography.
\newblock In \emph{Workshop on the Theory and Application of Cryptographic
  Techniques}, pages 475--480. Springer, 1984.
\newblock \doi{10.1007/3-540-39568-7_39}.

\bibitem[Bennett et~al.(1992)Bennett, Bessette, Brassard, Salvail, and
  Smolin]{Bennett1992}
Charles~H. Bennett, François Bessette, Gilles Brassard, Louis Salvail, and
  John Smolin.
\newblock Experimental quantum cryptography.
\newblock \emph{Journal of Cryptology}, 5:\penalty0 3--28, 1992.
\newblock \doi{10.1007/BF00191318}.

\bibitem[Jennewein et~al.(2000)Jennewein, Simon, Weihs, Weinfurter, and
  Zeilinger]{Jennewein2000}
Thomas. Jennewein, C.~Simon, Gregor Weihs, Harald Weinfurter, and Anton
  Zeilinger.
\newblock Quantum cryprography using entangled photons.
\newblock \emph{Physical Review Letters}, 84:\penalty0 4729, 2000.
\newblock \doi{10.1103/PhysRevLett.84.4729}.

\bibitem[Broadbent et~al.(2009)Broadbent, Fitzsimons, and
  Kashefi]{Broadbent2009}
Anne Broadbent, Joseph Fitzsimons, and Elham Kashefi.
\newblock Universal blind quantum computation.
\newblock In \emph{2009 50th Annual IEEE Symposium on Foundations of Computer
  Science}, pages 517--526. IEEE, 2009.
\newblock \doi{10.1109/FOCS.2009.36}.

\bibitem[Grover(1997)]{Grover1997}
Lov~K Grover.
\newblock Quantum telecomputation.
\newblock \emph{arXiv preprint quant-ph/9704012}, 1997.

\bibitem[Nickerson et~al.(2014)Nickerson, Fitzsimons, and
  Benjamin]{Nickerson2014}
Naomi~H Nickerson, Joseph~F Fitzsimons, and Simon~C Benjamin.
\newblock Freely scalable quantum technologies using cells of 5-to-50 qubits
  with very lossy and noisy photonic links.
\newblock \emph{Physical Review X}, 4\penalty0 (4):\penalty0 041041, 2014.
\newblock \doi{10.1103/PhysRevX.4.041041}.

\bibitem[Komar et~al.(2014)Komar, Kessler, Bishof, Jiang, S{\o}rensen, Ye, and
  Lukin]{Komar2014}
Peter Komar, Eric~M Kessler, Michael Bishof, Liang Jiang, Anders~S S{\o}rensen,
  Jun Ye, and Mikhail~D Lukin.
\newblock A quantum network of clocks.
\newblock \emph{Nature Physics}, 10\penalty0 (8):\penalty0 582, 2014.
\newblock \doi{10.1038/nphys3000}.

\bibitem[Gottesman et~al.(2012)Gottesman, Jennewein, and Croke]{Gottesman2012}
Daniel Gottesman, Thomas Jennewein, and Sarah Croke.
\newblock Longer-baseline telescopes using quantum repeaters.
\newblock \emph{Physical Review Letters}, 109\penalty0 (7):\penalty0 070503,
  2012.
\newblock \doi{10.1103/PhysRevLett.109.070503}.

\bibitem[Wootters and Zurek(1982)]{Wootters1982}
W.~K. Wootters and W.~H. Zurek.
\newblock A single quantum cannot be cloned.
\newblock \emph{Nature}, 299:\penalty0 802--803, 1982.
\newblock \doi{10.1038/299802a0}.

\bibitem[Dieks(1982)]{Dieks1982}
D.~Dieks.
\newblock Communication by {EPR} devices.
\newblock \emph{Physics Letters A}, 92:\penalty0 271--272, 1982.
\newblock \doi{10.1016/0375-9601(82)90084-6}.

\bibitem[Briegel et~al.(1998)Briegel, D{\"u}r, Cirac, and Zoller]{Briegel1998}
H-J Briegel, Wolfgang D{\"u}r, Juan~I Cirac, and Peter Zoller.
\newblock Quantum repeaters: the role of imperfect local operations in quantum
  communication.
\newblock \emph{Physical Review Letters}, 81\penalty0 (26):\penalty0 5932,
  1998.
\newblock \doi{10.1103/PhysRevLett.81.5932}.

\bibitem[D{\"u}r et~al.(1999)D{\"u}r, Briegel, Cirac, and Zoller]{Dur1999}
W~D{\"u}r, H-J Briegel, JI~Cirac, and P~Zoller.
\newblock Quantum repeaters based on entanglement purification.
\newblock \emph{Physical Review A}, 59\penalty0 (1):\penalty0 169, 1999.
\newblock \doi{10.1103/PhysRevA.59.169}.

\bibitem[Muralidharan et~al.(2016)Muralidharan, Li, Kim, L{\"u}tkenhaus, Lukin,
  and Jiang]{Muralidharan2016}
Sreraman Muralidharan, Linshu Li, Jungsang Kim, Norbert L{\"u}tkenhaus,
  Mikhail~D Lukin, and Liang Jiang.
\newblock Optimal architectures for long distance quantum communication.
\newblock \emph{Scientific reports}, 6:\penalty0 20463, 2016.
\newblock \doi{10.1038/srep20463}.

\bibitem[Duan et~al.(2001)Duan, Lukin, Cirac, and Zoller]{Duan2001}
L.~M. Duan, M.~D. Lukin, J.~I. Cirac, and P.~Zoller.
\newblock Long-distance quantum communication with atomic ensembles and linear
  optics.
\newblock \emph{Nature}, 414\penalty0 (6862):\penalty0 413--418, November 2001.
\newblock ISSN 0028-0836.
\newblock \doi{10.1038/35106500}.

\bibitem[Childress et~al.(2006)Childress, Taylor, S{\o}rensen, and
  Lukin]{Childress2006}
Lilian Childress, JM~Taylor, Anders~S{\o}ndberg S{\o}rensen, and MD~Lukin.
\newblock Fault-tolerant quantum communication based on solid-state photon
  emitters.
\newblock \emph{Physical Review Letters}, 96\penalty0 (7):\penalty0 070504,
  2006.
\newblock \doi{10.1103/PhysRevLett.96.070504}.

\bibitem[Hartmann et~al.(2007)Hartmann, Kraus, Briegel, and
  D{\"u}r]{Hartmann2007}
Lorenz Hartmann, Barbara Kraus, H-J Briegel, and W~D{\"u}r.
\newblock Role of memory errors in quantum repeaters.
\newblock \emph{Physical Review A}, 75\penalty0 (3):\penalty0 032310, 2007.
\newblock \doi{10.1103/PhysRevA.75.032310}.

\bibitem[Collins et~al.(2007)Collins, Jenkins, Kuzmich, and
  Kennedy]{Collins2007}
OA~Collins, SD~Jenkins, A~Kuzmich, and TAB Kennedy.
\newblock Multiplexed memory-insensitive quantum repeaters.
\newblock \emph{Physical review letters}, 98\penalty0 (6):\penalty0 060502,
  2007.
\newblock \doi{10.1103/PhysRevLett.98.060502}.

\bibitem[Sangouard et~al.(2011)Sangouard, Simon, De~Riedmatten, and
  Gisin]{Sangouard2011}
Nicolas Sangouard, Christoph Simon, Hugues De~Riedmatten, and Nicolas Gisin.
\newblock Quantum repeaters based on atomic ensembles and linear optics.
\newblock \emph{Reviews of Modern Physics}, 83\penalty0 (1):\penalty0 33, 2011.
\newblock \doi{10.1103/RevModPhys.83.33}.

\bibitem[Vinay and Kok(2017)]{Vinay2017}
Scott~E Vinay and Pieter Kok.
\newblock Practical repeaters for ultralong-distance quantum communication.
\newblock \emph{Physical Review A}, 95\penalty0 (5):\penalty0 052336, 2017.
\newblock \doi{10.1103/PhysRevA.95.052336}.

\bibitem[Rozp{\k{e}}dek et~al.(2019)Rozp{\k{e}}dek, Yehia, Goodenough, Ruf,
  Humphreys, Hanson, Wehner, and Elkouss]{Rozpcedek2019}
Filip Rozp{\k{e}}dek, Raja Yehia, Kenneth Goodenough, Maximilian Ruf, Peter~C
  Humphreys, Ronald Hanson, Stephanie Wehner, and David Elkouss.
\newblock Near-term quantum-repeater experiments with nitrogen-vacancy centers:
  Overcoming the limitations of direct transmission.
\newblock \emph{Physical Review A}, 99\penalty0 (5):\penalty0 052330, 2019.
\newblock \doi{10.1103/PhysRevA.99.052330}.

\bibitem[Bhaskar et~al.(2020)Bhaskar, Riedinger, Machielse, Levonian, Nguyen,
  Knall, Park, Englund, Lon{\v{c}}ar, Sukachev, et~al.]{Bhaskar2020}
Mihir~K Bhaskar, Ralf Riedinger, Bartholomeus Machielse, David~S Levonian,
  Christian~T Nguyen, Erik~N Knall, Hongkun Park, Dirk Englund, Marko
  Lon{\v{c}}ar, Denis~D Sukachev, et~al.
\newblock Experimental demonstration of memory-enhanced quantum communication.
\newblock \emph{Nature}, 580\penalty0 (7801):\penalty0 60--64, 2020.
\newblock \doi{10.1038/s41586-020-2103-5}.

\bibitem[Khatri et~al.(2019)Khatri, Matyas, Siddiqui, and Dowling]{Khatri2019}
Sumeet Khatri, Corey~T Matyas, Aliza~U Siddiqui, and Jonathan~P Dowling.
\newblock Practical figures of merit and thresholds for entanglement
  distribution in quantum networks.
\newblock \emph{Physical Review Research}, 1\penalty0 (2):\penalty0 023032,
  2019.
\newblock \doi{10.1103/PhysRevResearch.1.023032}.

\bibitem[Cabrillo et~al.(1999)Cabrillo, Cirac, Garcia-Fernandez, and
  Zoller]{Cabrillo1999}
C~Cabrillo, JI~Cirac, P~Garcia-Fernandez, and P~Zoller.
\newblock Creation of entangled states of distant atoms by interference.
\newblock \emph{Physical Review A}, 59\penalty0 (2):\penalty0 1025, 1999.
\newblock \doi{10.1103/PhysRevA.59.1025}.

\bibitem[Barrett and Kok(2005)]{Barrett2005}
Sean~D Barrett and Pieter Kok.
\newblock Efficient high-fidelity quantum computation using matter qubits and
  linear optics.
\newblock \emph{Physical Review A}, 71\penalty0 (6):\penalty0 060310, 2005.
\newblock \doi{10.1103/PhysRevA.71.060310}.

\bibitem[Duan and Kimble(2004)]{Duan2004}
L.-M. Duan and H.~J. Kimble.
\newblock Scalable photonic quantum computation through cavity-assisted
  interactions.
\newblock \emph{Physical Review Letters}, 92\penalty0 (12):\penalty0 127902,
  2004.
\newblock \doi{10.1103/PhysRevLett.92.127902}.

\bibitem[Jiang et~al.(2009)Jiang, Taylor, Nemoto, Munro, Van~Meter, and
  Lukin]{Jiang2009}
Liang Jiang, Jacob~M Taylor, Kae Nemoto, William~J Munro, Rodney Van~Meter, and
  Mikhail~D Lukin.
\newblock Quantum repeater with encoding.
\newblock \emph{Physical Review A}, 79\penalty0 (3):\penalty0 032325, 2009.
\newblock \doi{10.1103/PhysRevA.79.032325}.

\bibitem[Munro et~al.(2012)Munro, Stephens, Devitt, Harrison, and
  Nemoto]{Munro2012}
William~J Munro, Ashley~M Stephens, Simon~J Devitt, Keith~A Harrison, and Kae
  Nemoto.
\newblock Quantum communication without the necessity of quantum memories.
\newblock \emph{Nature Photonics}, 6\penalty0 (11):\penalty0 777, 2012.
\newblock \doi{10.1038/nphoton.2012.243}.

\bibitem[Muralidharan et~al.(2014)Muralidharan, Kim, L{\"u}tkenhaus, Lukin, and
  Jiang]{Muralidharan2014}
Sreraman Muralidharan, Jungsang Kim, Norbert L{\"u}tkenhaus, Mikhail~D Lukin,
  and Liang Jiang.
\newblock Ultrafast and fault-tolerant quantum communication across long
  distances.
\newblock \emph{Physical review letters}, 112\penalty0 (25):\penalty0 250501,
  2014.
\newblock \doi{10.1103/PhysRevLett.112.250501}.

\bibitem[Azuma et~al.(2015)Azuma, Tamaki, and Lo]{Azuma2015}
Koji Azuma, Kiyoshi Tamaki, and Hoi-Kwong Lo.
\newblock All-photonic quantum repeaters.
\newblock \emph{Nature communications}, 6:\penalty0 6787, 2015.
\newblock \doi{10.1038/ncomms7787}.

\bibitem[Ewert et~al.(2016)Ewert, Bergmann, and van Loock]{Ewert2016}
Fabian Ewert, Marcel Bergmann, and Peter van Loock.
\newblock Ultrafast long-distance quantum communication with static linear
  optics.
\newblock \emph{Physical review letters}, 117\penalty0 (21):\penalty0 210501,
  2016.
\newblock \doi{10.1103/PhysRevLett.117.210501}.

\bibitem[Hasegawa et~al.(2019)Hasegawa, Ikuta, Matsuda, Tamaki, Lo, Yamamoto,
  Azuma, and Imoto]{Hasegawa2019}
Yasushi Hasegawa, Rikizo Ikuta, Nobuyuki Matsuda, Kiyoshi Tamaki, Hoi-Kwong Lo,
  Takashi Yamamoto, Koji Azuma, and Nobuyuki Imoto.
\newblock Experimental time-reversed adaptive bell measurement towards
  all-photonic quantum repeaters.
\newblock \emph{Nature communications}, 10\penalty0 (1):\penalty0 378, 2019.
\newblock \doi{10.1038/s41467-018-08099-5}.

\bibitem[Li et~al.(2019)Li, Zhang, Yin, Liu, Hu, Fang, Fei, Jiang, Zhang, Li,
  et~al.]{Li2019}
Zheng-Da Li, Rui Zhang, Xu-Fei Yin, Li-Zheng Liu, Yi~Hu, Yu-Qiang Fang,
  Yue-Yang Fei, Xiao Jiang, Jun Zhang, Li~Li, et~al.
\newblock Experimental quantum repeater without quantum memory.
\newblock \emph{Nature Photonics}, page~1, 2019.
\newblock \doi{10.1038/s41566-019-0468-5}.

\bibitem[Varnava et~al.(2006)Varnava, Browne, and Rudolph]{Varnava2006}
Michael Varnava, Daniel~E Browne, and Terry Rudolph.
\newblock Loss tolerance in one-way quantum computation via counterfactual
  error correction.
\newblock \emph{Physical review letters}, 97\penalty0 (12):\penalty0 120501,
  2006.
\newblock \doi{10.1103/PhysRevLett.97.120501}.

\bibitem[Browne and Rudolph(2005)]{Browne2005}
Daniel~E Browne and Terry Rudolph.
\newblock Resource-efficient linear optical quantum computation.
\newblock \emph{Physical Review Letters}, 95\penalty0 (1):\penalty0 010501,
  2005.
\newblock \doi{10.1103/PhysRevLett.95.010501}.

\bibitem[Pant et~al.(2017)Pant, Krovi, Englund, and Guha]{Pant2017}
Mihir Pant, Hari Krovi, Dirk Englund, and Saikat Guha.
\newblock Rate-distance tradeoff and resource costs for all-optical quantum
  repeaters.
\newblock \emph{Physical Review A}, 95\penalty0 (1):\penalty0 012304, 2017.
\newblock \doi{10.1103/PhysRevA.95.012304}.

\bibitem[Lindner and Rudolph(2009)]{Lindner2009}
Netanel~H Lindner and Terry Rudolph.
\newblock Proposal for pulsed on-demand sources of photonic cluster state
  strings.
\newblock \emph{Physical Review Letters}, 103\penalty0 (11):\penalty0 113602,
  2009.
\newblock \doi{10.1103/PhysRevLett.103.113602}.

\bibitem[Saavedra et~al.(2000)Saavedra, Gheri, T{\"o}rm{\"a}, Cirac, and
  Zoller]{Saavedra2000}
C~Saavedra, KM~Gheri, P~T{\"o}rm{\"a}, JI~Cirac, and P~Zoller.
\newblock Controlled source of entangled photonic qubits.
\newblock \emph{Physical Review A}, 61\penalty0 (6):\penalty0 062311, 2000.
\newblock \doi{10.1103/PhysRevA.61.062311}.

\bibitem[Schwartz et~al.(2016)Schwartz, Cogan, Schmidgall, Don, Gantz, Kenneth,
  Lindner, and Gershoni]{Schwartz2016}
Ido Schwartz, Dan Cogan, Emma~R Schmidgall, Yaroslav Don, Liron Gantz, Oded
  Kenneth, Netanel~H Lindner, and David Gershoni.
\newblock Deterministic generation of a cluster state of entangled photons.
\newblock \emph{Science}, 354:\penalty0 434--437, 2016.
\newblock \doi{10.1126/science.aah4758}.

\bibitem[Economou et~al.(2010)Economou, Lindner, and Rudolph]{Economou2010}
Sophia~E Economou, Netanel Lindner, and Terry Rudolph.
\newblock Optically generated 2-dimensional photonic cluster state from coupled
  quantum dots.
\newblock \emph{Physical review letters}, 105\penalty0 (9):\penalty0 093601,
  2010.
\newblock \doi{10.1103/PhysRevLett.105.093601}.

\bibitem[Buterakos et~al.(2017)Buterakos, Barnes, and Economou]{Buterakos2017}
Donovan Buterakos, Edwin Barnes, and Sophia~E Economou.
\newblock Deterministic generation of all-photonic quantum repeaters from
  solid-state emitters.
\newblock \emph{Physical Review X}, 7\penalty0 (4):\penalty0 041023, 2017.
\newblock \doi{10.1103/PhysRevX.7.041023}.

\bibitem[Russo et~al.(2019)Russo, Barnes, and Economou]{Russo2019}
Antonio Russo, Edwin Barnes, and Sophia~E Economou.
\newblock Generation of arbitrary all-photonic graph states from quantum
  emitters.
\newblock \emph{New Journal of Physics}, 21\penalty0 (5):\penalty0 055002,
  2019.
\newblock \doi{10.1088/1367-2630/ab193d}.

\bibitem[Pichler et~al.(2017)Pichler, Choi, Zoller, and Lukin]{Pichler2017}
Hannes Pichler, Soonwon Choi, Peter Zoller, and Mikhail~D Lukin.
\newblock Universal photonic quantum computation via time-delayed feedback.
\newblock \emph{Proceedings of the National Academy of Sciences}, 114\penalty0
  (43):\penalty0 11362--11367, 2017.
\newblock \doi{10.1073/pnas.1711003114}.

\bibitem[Russo et~al.(2018)Russo, Barnes, and Economou]{Russo2018}
Antonio Russo, Edwin Barnes, and Sophia~E Economou.
\newblock Photonic graph state generation from quantum dots and color centers
  for quantum communications.
\newblock \emph{Physical Review B}, 98\penalty0 (8):\penalty0 085303, 2018.
\newblock \doi{10.1103/PhysRevB.98.085303}.

\bibitem[Chan(2018)]{Chan2018}
Ming~Lai Chan.
\newblock Optimized protocol to create repeater graph states for all-photonic
  quantum repeater.
\newblock \emph{arXiv preprint arXiv:1811.10214}, 2018.

\bibitem[Gimeno-Segovia et~al.(2019)Gimeno-Segovia, Rudolph, and
  Economou]{Gimeno2019}
Mercedes Gimeno-Segovia, Terry Rudolph, and Sophia~E Economou.
\newblock Deterministic generation of large-scale entangled photonic cluster
  state from interacting solid state emitters.
\newblock \emph{Physical review letters}, 123\penalty0 (7):\penalty0 070501,
  2019.
\newblock \doi{10.1103/PhysRevLett.123.070501}.

\bibitem[Borregaard et~al.(2020)Borregaard, Pichler, Schr{\"o}der, Lukin,
  Lodahl, and S{\o}rensen]{Borregaard2020}
Johannes Borregaard, Hannes Pichler, Tim Schr{\"o}der, Mikhail~D Lukin, Peter
  Lodahl, and Anders~S S{\o}rensen.
\newblock One-way quantum repeater based on near-deterministic photon-emitter
  interfaces.
\newblock \emph{Physical Review X}, 10\penalty0 (2):\penalty0 021071, 2020.
\newblock \doi{10.1103/PhysRevX.10.021071}.

\bibitem[Lloyd et~al.(2001)Lloyd, Shahriar, Shapiro, and Hemmer]{Lloyd2001}
S~Lloyd, MS~Shahriar, JH~Shapiro, and PR~Hemmer.
\newblock Long distance, unconditional teleportation of atomic states via
  complete bell state measurements.
\newblock \emph{Physical Review Letters}, 87\penalty0 (16):\penalty0 167903,
  2001.
\newblock \doi{10.1103/PhysRevLett.87.167903}.

\bibitem[Kim et~al.(2001)Kim, Kulik, and Shih]{Kim2001}
Yoon-Ho Kim, Sergei~P Kulik, and Yanhua Shih.
\newblock Quantum teleportation of a polarization state with a complete bell
  state measurement.
\newblock \emph{Physical Review Letters}, 86\penalty0 (7):\penalty0 1370, 2001.
\newblock \doi{10.1103/PhysRevLett.86.1370}.

\bibitem[Kim et~al.(2002)Kim, KULIK, and Shih]{Kim2002}
Yoon-Ho Kim, SERGEI KULIK, and Yanhua Shih.
\newblock Quantum teleportation with a complete bell state measurement.
\newblock \emph{Journal of Modern Optics}, 49\penalty0 (1-2):\penalty0
  221--236, 2002.
\newblock \doi{10.1080/09500340110087633}.

\bibitem[Grice(2011)]{Grice2011}
Warren~P Grice.
\newblock Arbitrarily complete bell-state measurement using only linear optical
  elements.
\newblock \emph{Physical Review A}, 84\penalty0 (4):\penalty0 042331, 2011.
\newblock \doi{10.1103/PhysRevA.84.042331}.

\bibitem[Ewert and van Loock(2014)]{Ewert2014}
Fabian Ewert and Peter van Loock.
\newblock 3/4-efficient bell measurement with passive linear optics and
  unentangled ancillae.
\newblock \emph{Physical review letters}, 113\penalty0 (14):\penalty0 140403,
  2014.
\newblock \doi{10.1103/PhysRevLett.113.140403}.

\bibitem[Wein et~al.(2016)Wein, Heshami, Fuchs, Krovi, Dutton, Tittel, and
  Simon]{Wein2016}
Stephen Wein, Khabat Heshami, Christopher~A Fuchs, Hari Krovi, Zachary Dutton,
  Wolfgang Tittel, and Christoph Simon.
\newblock Efficiency of an enhanced linear optical bell-state measurement
  scheme with realistic imperfections.
\newblock \emph{Physical Review A}, 94\penalty0 (3):\penalty0 032332, 2016.
\newblock \doi{10.1103/PhysRevA.94.032332}.

\bibitem[Martin and Whaley(2019)]{Martin2019}
Leigh~S Martin and K~Birgitta Whaley.
\newblock Single-shot deterministic entanglement between non-interacting
  systems with linear optics.
\newblock \emph{arXiv preprint arXiv:1912.00067}, 2019.

\bibitem[Hein et~al.(2004)Hein, Eisert, and Briegel]{Hein2004}
Marc Hein, Jens Eisert, and Hans~J Briegel.
\newblock Multiparty entanglement in graph states.
\newblock \emph{Physical Review A}, 69\penalty0 (6):\penalty0 062311, 2004.
\newblock \doi{10.1103/PhysRevA.69.062311}.

\bibitem[Takeoka et~al.(2014)Takeoka, Guha, and Wilde]{Takeoka2014}
Masahiro Takeoka, Saikat Guha, and Mark~M Wilde.
\newblock Fundamental rate-loss tradeoff for optical quantum key distribution.
\newblock \emph{Nature communications}, 5\penalty0 (1):\penalty0 1--7, 2014.
\newblock \doi{10.1038/ncomms6235}.

\bibitem[Pirandola et~al.(2017)Pirandola, Laurenza, Ottaviani, and
  Banchi]{Pirandola2017}
Stefano Pirandola, Riccardo Laurenza, Carlo Ottaviani, and Leonardo Banchi.
\newblock Fundamental limits of repeaterless quantum communications.
\newblock \emph{Nature communications}, 8\penalty0 (1):\penalty0 1--15, 2017.
\newblock \doi{10.1038/ncomms15043}.

\bibitem[Shor and Preskill(2000)]{Shor2000}
Peter~W Shor and John Preskill.
\newblock Simple proof of security of the bb84 quantum key distribution
  protocol.
\newblock \emph{Physical review letters}, 85\penalty0 (2):\penalty0 441, 2000.
\newblock \doi{10.1103/PhysRevLett.85.441}.

\bibitem[Scarani et~al.(2009)Scarani, Bechmann-Pasquinucci, Cerf, Du{\v{s}}ek,
  L{\"u}tkenhaus, and Peev]{Scarani2009}
Valerio Scarani, Helle Bechmann-Pasquinucci, Nicolas~J Cerf, Miloslav
  Du{\v{s}}ek, Norbert L{\"u}tkenhaus, and Momtchil Peev.
\newblock The security of practical quantum key distribution.
\newblock \emph{Reviews of modern physics}, 81\penalty0 (3):\penalty0 1301,
  2009.
\newblock \doi{10.1103/RevModPhys.81.1301}.

\bibitem[Pan et~al.(2001)Pan, Simon, Brukner, and Zeilinger]{Pan2001}
Jian-Wei Pan, Christoph Simon, {\v{C}}aslav Brukner, and Anton Zeilinger.
\newblock Entanglement purification for quantum communication.
\newblock \emph{Nature}, 410\penalty0 (6832):\penalty0 1067, 2001.
\newblock \doi{10.1038/35074041}.

\bibitem[Zwerger et~al.(2012)Zwerger, D{\"u}r, and Briegel]{Zwerger2012}
M~Zwerger, W~D{\"u}r, and HJ~Briegel.
\newblock Measurement-based quantum repeaters.
\newblock \emph{Physical Review A}, 85\penalty0 (6):\penalty0 062326, 2012.
\newblock \doi{10.1103/PhysRevA.85.062326}.

\bibitem[Zwerger et~al.(2016)Zwerger, Briegel, and D{\"u}r]{Zwerger2016}
M~Zwerger, HJ~Briegel, and W~D{\"u}r.
\newblock Measurement-based quantum communication.
\newblock \emph{Applied Physics B}, 122\penalty0 (3):\penalty0 50, 2016.
\newblock \doi{10.1007/s00340-015-6285-8}.

\bibitem[Greilich et~al.(2011)Greilich, Carter, Kim, Bracker, and
  Gammon]{Greilich2011}
Alex Greilich, Samuel~G. Carter, Danny Kim, Allan~S. Bracker, and Daniel
  Gammon.
\newblock Optical control of one and two hole spins in interacting quantum
  dots.
\newblock \emph{Nature Photonics}, 5\penalty0 (11):\penalty0 702--708, November
  2011.
\newblock \doi{10.1038/nphoton.2011.237}.

\bibitem[Kim et~al.(2011)Kim, Carter, Greilich, Bracker, and Gammon]{Kim2011}
Danny Kim, Samuel~G Carter, Alex Greilich, Allan~S Bracker, and Daniel Gammon.
\newblock Ultrafast optical control of entanglement between two quantum-dot
  spins.
\newblock \emph{Nature Physics}, 7\penalty0 (3):\penalty0 223--229, 2011.
\newblock \doi{10.1038/nphys1863}.

\bibitem[Greilich et~al.(2007)Greilich, Shabaev, Yakovlev, Efros, Yugova,
  Reuter, Wieck, and Bayer]{Greilich2007}
A.~Greilich, A.~Shabaev, D.~R. Yakovlev, Al.~L. Efros, I.~A. Yugova, D.~Reuter,
  A.~D. Wieck, and M.~Bayer.
\newblock Nuclei-induced frequency focusing of electron spin coherence.
\newblock \emph{Science}, 317\penalty0 (5846):\penalty0 1896--1899, 2007.
\newblock \doi{10.1126/science.1146850}.

\bibitem[Wang et~al.(2012)Wang, Chesi, and Coish]{Wang2012}
Xiaoya~Judy Wang, Stefano Chesi, and W.~A. Coish.
\newblock Spin-echo dynamics of a heavy hole in a quantum dot.
\newblock \emph{Phys. Rev. Lett.}, 109:\penalty0 237601, 2012.
\newblock \doi{10.1103/PhysRevLett.109.237601}.

\bibitem[Najer et~al.(2019)Najer, S{\"o}llner, Sekatski, Dolique, L{\"o}bl,
  Riedel, Schott, Starosielec, Valentin, Wieck, et~al.]{Najer2019}
Daniel Najer, Immo S{\"o}llner, Pavel Sekatski, Vincent Dolique, Matthias~C
  L{\"o}bl, Daniel Riedel, R{\"u}diger Schott, Sebastian Starosielec, Sascha~R
  Valentin, Andreas~D Wieck, et~al.
\newblock A gated quantum dot strongly coupled to an optical microcavity.
\newblock \emph{Nature}, pages 1--1, 2019.
\newblock \doi{10.1038/s41586-019-1709-y}.

\bibitem[Fuchs et~al.(2011)Fuchs, Burkard, Klimov, and Awschalom]{Fuchs2011}
GD~Fuchs, Guido Burkard, PV~Klimov, and DD~Awschalom.
\newblock A quantum memory intrinsic to single nitrogen--vacancy centres in
  diamond.
\newblock \emph{Nature Physics}, 7\penalty0 (10):\penalty0 789--793, 2011.
\newblock \doi{10.1038/nphys2026}.

\bibitem[Solenov et~al.(2013)Solenov, Economou, and Reinecke]{Solenov2013}
Dmitry Solenov, Sophia~E Economou, and Thomas~L Reinecke.
\newblock Two-qubit quantum gates for defect qubits in diamond and similar
  systems.
\newblock \emph{Physical Review B}, 88\penalty0 (16):\penalty0 161403, 2013.
\newblock \doi{10.1103/PhysRevB.88.161403}.

\bibitem[Goldman et~al.(2020)Goldman, Patti, Levonian, Yelin, and
  Lukin]{Goldman2020}
ML~Goldman, TL~Patti, D~Levonian, SF~Yelin, and MD~Lukin.
\newblock Optical control of a single nuclear spin in the solid state.
\newblock \emph{Physical Review Letters}, 124\penalty0 (15):\penalty0 153203,
  2020.
\newblock \doi{10.1103/PhysRevLett.124.153203}.

\bibitem[Rong et~al.(2015)Rong, Geng, Shi, Liu, Xu, Ma, Kong, Jiang, Wu, and
  Du]{Rong2015}
Xing Rong, Jianpei Geng, Fazhan Shi, Ying Liu, Kebiao Xu, Wenchao Ma, Fei Kong,
  Zhen Jiang, Yang Wu, and Jiangfeng Du.
\newblock Experimental fault-tolerant universal quantum gates with solid-state
  spins under ambient conditions.
\newblock \emph{Nature communications}, 6\penalty0 (1):\penalty0 1--7, 2015.
\newblock \doi{10.1038/ncomms9748 (2015)}.

\bibitem[Bar-Gill et~al.(2013)Bar-Gill, Pham, Jarmola, Budker, and
  Walsworth]{Bar-Gill2013}
Nir Bar-Gill, Linh~M Pham, Andrejs Jarmola, Dmitry Budker, and Ronald~L
  Walsworth.
\newblock Solid-state electronic spin coherence time approaching one second.
\newblock \emph{Nature communications}, 4:\penalty0 1743, 2013.
\newblock \doi{10.1038/ncomms2771}.

\bibitem[Ballance et~al.(2016)Ballance, Harty, Linke, Sepiol, and
  Lucas]{Ballance2016}
CJ~Ballance, TP~Harty, NM~Linke, MA~Sepiol, and DM~Lucas.
\newblock High-fidelity quantum logic gates using trapped-ion hyperfine qubits.
\newblock \emph{Physical review letters}, 117\penalty0 (6):\penalty0 060504,
  2016.
\newblock \doi{10.1103/PhysRevLett.117.060504}.

\bibitem[Wang et~al.(2017)Wang, Um, Zhang, An, Lyu, Zhang, Duan, Yum, and
  Kim]{Wang2017b}
Ye~Wang, Mark Um, Junhua Zhang, Shuoming An, Ming Lyu, Jing-Ning Zhang, L-M
  Duan, Dahyun Yum, and Kihwan Kim.
\newblock Single-qubit quantum memory exceeding ten-minute coherence time.
\newblock \emph{Nature Photonics}, 11\penalty0 (10):\penalty0 646, 2017.
\newblock \doi{10.1038/s41566-017-0007-1}.

\bibitem[Daiss et~al.(2019)Daiss, Welte, Hacker, Li, and Rempe]{Daiss2019}
Severin Daiss, Stephan Welte, Bastian Hacker, Lin Li, and Gerhard Rempe.
\newblock Single-photon distillation via a photonic parity measurement using
  cavity qed.
\newblock \emph{Physical review letters}, 122\penalty0 (13):\penalty0 133603,
  2019.
\newblock \doi{10.1103/PhysRevLett.122.133603}.

\bibitem[Hu et~al.(2008{\natexlab{a}})Hu, Young, O'Brien, Munro, and
  Rarity]{Hu2008}
C.~Y. Hu, A.~Young, J.~L. O'Brien, W.~J. Munro, and J.~G. Rarity.
\newblock Giant optical {Faraday} rotation induced by a single-electron spin in
  a quantum dot: Applications to entangling remote spins via a single photon.
\newblock \emph{Physical Review B}, 78\penalty0 (8):\penalty0 085307, Aug
  2008{\natexlab{a}}.
\newblock \doi{10.1103/PhysRevB.78.085307}.
\newblock URL \url{http://prb.aps.org/abstract/PRB/v78/i8/e085307}.

\bibitem[Koshino et~al.(2010)Koshino, Ishizaka, and Nakamura]{Koshino2010}
Kazuki Koshino, Satoshi Ishizaka, and Yasunobu Nakamura.
\newblock Deterministic photon-photon swap gate using a $\lambda$ system.
\newblock \emph{Physical Review A}, 82\penalty0 (1):\penalty0 010301, 2010.
\newblock \doi{10.1103/PhysRevA.82.010301}.

\bibitem[Rosenblum et~al.(2011)Rosenblum, Parkins, and Dayan]{Rosenblum2011}
Serge Rosenblum, Scott Parkins, and Barak Dayan.
\newblock Photon routing in cavity {QED}: Beyond the fundamental limit of
  photon blockade.
\newblock \emph{Physical Review A}, 84:\penalty0 033854, 2011.
\newblock \doi{10.1103/PhysRevA.84.033854}.

\bibitem[Sun et~al.(2016)Sun, Kim, Solomon, and Waks]{Sun2016}
Shuo Sun, Hyochul Kim, Glenn~S Solomon, and Edo Waks.
\newblock A quantum phase switch between a single solid-state spin and a
  photon.
\newblock \emph{Nature Nanotechnology}, 11\penalty0 (6):\penalty0 539--544,
  2016.
\newblock \doi{10.1038/nnano.2015.334}.

\bibitem[Wells et~al.(2019)Wells, Kalliakos, Villa, Ellis, Stevenson, Bennett,
  Farrer, Ritchie, and Shields]{Wells2019}
LM~Wells, Sokratis Kalliakos, Bruno Villa, DJP Ellis, RM~Stevenson, AJ~Bennett,
  Ian Farrer, DA~Ritchie, and AJ~Shields.
\newblock Photon phase shift at the few-photon level and optical switching by a
  quantum dot in a microcavity.
\newblock \emph{Physical Review Applied}, 11\penalty0 (6):\penalty0 061001,
  2019.
\newblock \doi{10.1103/PhysRevApplied.11.061001}.

\bibitem[Androvitsaneas et~al.(2019)Androvitsaneas, Young, Lennon, Schneider,
  Maier, Hinchliff, Atkinson, Harbord, Kamp, H{\"o}fling,
  et~al.]{Androvitsaneas2019}
Petros Androvitsaneas, Andrew Young, Joseph Lennon, Christian Schneider,
  Sebastian Maier, Janna Hinchliff, George Atkinson, Edmund Harbord, Martin
  Kamp, Sven H{\"o}fling, et~al.
\newblock An efficient quantum photonic phase shift in a low q-factor regime.
\newblock \emph{ACS Photonics}, 2019.
\newblock \doi{10.1021/acsphotonics.8b01380}.

\bibitem[Hadfield(2009)]{Hadfield2009}
Robert~H Hadfield.
\newblock Single-photon detectors for optical quantum information applications.
\newblock \emph{Nature photonics}, 3\penalty0 (12):\penalty0 696, 2009.
\newblock \doi{10.1038/nphoton.2009.230}.

\bibitem[Zhan and Sun(2020)]{Zhan2020}
Yuan Zhan and Shuo Sun.
\newblock Deterministic generation of loss-tolerant photonic cluster states
  with a single quantum emitter.
\newblock \emph{Physical Review Letters}, 125\penalty0 (22):\penalty0 223601,
  2020.
\newblock \doi{10.1103/PhysRevLett.125.223601}.

\bibitem[Hu et~al.(2008{\natexlab{b}})Hu, Young, O'Brien, Munro, and
  Rarity]{Hu2008a}
C.~Y. Hu, A.~Young, J.~L. O'Brien, W.~J. Munro, and J.~G. Rarity.
\newblock Giant optical {Faraday} rotation induced by a single-electron spin in
  a quantum dot: Applications to entangling remote spins via a single photon.
\newblock \emph{Physical Review B}, 78\penalty0 (8):\penalty0 085307,
  2008{\natexlab{b}}.
\newblock \doi{10.1103/PhysRevB.78.085307}.

\end{thebibliography}

\newpage
\clearpage

\onecolumn

\title{\textbf{\Large Supplementary Materials: Resource requirements for efficient quantum communication using all-photonic graph states generated from a few matter qubits}}

\maketitle

\section*{Bounds for memory-based repeater protocols}
As discussed in the main text, the rate of a memory-based quantum repeater based on heralded entanglement generation is limited by the rate $\langle T_{QM} \rangle^{-1}$ at which a given memory qubit can be re-used. $\langle T_{QM} \rangle$ is given by the sum of the average time required for generating entanglement $\langle T_{\mathrm{ent}} \rangle$ and the time during which the entanglement is stored $\langle T_{\mathrm{store}} \rangle$. $\langle T_{\mathrm{ent}} \rangle$ is given by $T_{\mathrm{trial}}/P_{\mathrm{ent}}$, where $T_{\mathrm{trial}}$ is the time required for a heralded entanglement generation attempt, and $P_{\mathrm{ent}}$ is the probability that an attempt is successful. Let's consider three heralded entanglement procedures~\cite{Cabrillo1999, Barrett2005, Duan2004}:
\begin{itemize}
    \item In Ref.~\cite{Barrett2005}, $T_{\mathrm{trial}} = L_0/c$, and the probability is $P_{\mathrm{ent}} = e^{-L_0/L_{\mathrm{att}}} / 2$ (assuming perfect photon emission and detection probability).
    \item In Ref.~\cite{Duan2004}, $T_{\mathrm{trial}} = 2L_0/c$, and the probability is $P_{\mathrm{ent}} = e^{-L_0/L_{\mathrm{att}}}$.
    \item In Ref.~\cite{Cabrillo1999}, $T_{\mathrm{trial}} = L_0/c$, and the probability is $P_{\mathrm{ent}} \propto e^{-L_0/(2 L_{\mathrm{att}})}$. However, the protocol needs to be implemented in a regime where the probability to emit a photon is very small to avoid two-photon emission, which degrades the entanglement fidelity. Consequently, $P_{\mathrm{ent}}\ll 1/2$.
\end{itemize}
Thus in all these cases, if we consider small values of $L_0\ll L_{\mathrm{att}}$, we have:
\begin{equation}
    \langle T_{\mathrm{ent}} \rangle \geq \frac{2L_0}{c} e^{L_0/L_{\mathrm{att}}} \geq \frac{2L_0}{c}.
\end{equation}

This limitation can be understood from the fact that the photonic qubit and the classical signal need to travel back and forth either to a measurement node (twice the distance $L_0/2$) or to the next QR (twice the distance $L_0)$. If the photon travels to a quantum repeater, it is possible to deterministically entangle the QM and the photon, so the entanglement probability is only bounded by the photonic losses. If the photon travels to an intermediate node, the entanglement generation procedure uses linear optics and has a success probability that cannot reach more than $1/2$ (unless ancillary photonic qubits are used, but this requires another quantum emitter to produce them~\cite{Grice2011,  Ewert2014, Wein2016}).

From this limitation, we find the upper bound $R_{\mathrm{max}}^{(QM)} \geq \max(R/N_m)$ in the case of QR protocols with two QMs. In this case, the rate $R$ is limited by $\langle T_{QM} \rangle^{-1}\leq \langle T_{\mathrm{ent}} \rangle^{-1} = c / (2L_0)$. The number of matter qubits is given by twice the number of repeaters plus Alice and Bob's matter qubits, thus $N_m = 2 (N_{QR} + 1) = 2 L/L_0$, therefore:
\begin{equation}
    R_{\mathrm{max}}^{(QM)} = c / 4L.\label{eq:memoryBasedBound1}
\end{equation}

In the case of multiplexed repeaters, by multiplying the number of memories per repeater by $N$, the entangling rate between two repeaters is bounded by $N c/(2L_0)$, but the number of matter qubits is now $N_m=2N L/L_0$, so that the total rate is still bounded by $R_{\mathrm{max}}^{(QM)}$.

Ref.~\cite{Martin2019} also proposes a method to generate  heralded entanglement deterministically by detecting photons. The idea is to excite the two quantum emitters (each consisting of two ground states and one excited state that couples to only one of the ground states) such that they both emit a photon whose which-path information is erased using linear optics. When the first photon is detected a signal is sent to each quantum emitter to perform a rotation between the ground states. This rotation needs to be performed before the second photon is emitted. Therefore, the latency between the photon detection and the rotation should be much smaller than the lifetime of the quantum emitter's excited state. A direct implication of this is that the photon detectors should be placed very close to the quantum emitters, which is impractical. For example, for quantum emitters with a lifetime under $10 {\rm \; ns}$ (which is typical for QDs and defects in diamond), the position of the detectors should be much smaller than $1 {\rm \; m}$.
In addition, to obtain high fidelities with this scheme, the two quantum emitters are excited for a typical time $T_1$ which should be much larger than the classical signaling time $L_0/c$, and thus the previous analysis should still hold in that case.

In the case of memory-based schemes using heralded entanglement and 2 QMs per QR, it is possible to find a stricter upper bound.
To derive this bound, consider two repeater nodes labeled 1 and 2. Node 1 contains two memory qubits QM1 and QM1$'$, while node 2 contains QM2 and QM2$'$. Suppose that we succeed in entangling QM1 and QM2 via a Bell measurement at an intermediate measurement node. How long do we have to maintain this entanglement before further entanglement swappings at neighboring nodes free up QM1 and QM2, allowing them to be re-used for creating the next Bell pair to be shared between Alice and Bob? The answer to this question is what we called $\langle T_{\rm store}\rangle$ above, the average amount of time we have to store the entanglement. The upper bound on the rate is then $\langle T_{\rm QM}\rangle^{-1}$, where $\langle T_{\rm QM}\rangle=\langle T_{\rm ent}\rangle+\langle T_{\rm store}\rangle$. We obtained a bound for $\langle T_{\rm ent}\rangle$ above and used this to derive the bound on the rate given in Eq.~\eqref{eq:memoryBasedBound1}. There, we neglected the role of $\langle T_{\rm store}\rangle$. Here, we reinstate this quantity to obtain a stricter bound.

To determine $\langle T_{\rm store}\rangle$, we first note that the entanglement between QM1 and QM2 needs to be maintained until QM1$'$ and QM2$'$ become entangled with memory qubits on other nodes. Once this happens, the entanglement can then be swapped, leaving QM1 and QM2 completely disentangled and ready for re-use. Of course, QM1$'$ and QM2$'$ could already be entangled with other nodes by the time QM1 and QM2 become entangled. Thus, there are four equally probable cases to consider:
\begin{itemize}
\item If QM1$'$ and QM2$'$ are already connected to other nodes, we immediately swap the entanglement and thus do not wait any longer. In this case $\langle T_{\mathrm{store}}^{\rm case 1}\rangle = 0$.
\item If one QM (either QM1$'$ or QM2$'$) is already connected to other nodes, we need to wait an average time of $\langle T_{\mathrm{ent}} \rangle$ for the second QM to become entangled as well. In this case, $\langle T_{\rm store}^{\rm case 2}\rangle=\langle T_{\rm store}^{\rm case 3}\rangle= \langle T_{\mathrm{ent}} \rangle$.
\item If neither QM1$'$ or QM2$'$ are connected, we need to wait for both to become connected. In this case, $\langle T_{\mathrm{store}}^{\rm case 4}\rangle = \frac{3-2P_{\mathrm{ent}}}{2- P_{\mathrm{ent}}}\times \langle T_{\mathrm{ent}} \rangle $.
\end{itemize}
The expression for $\langle T_{\mathrm{store}}^{\rm case 4}\rangle$ above requires some explanation. It comes from the fact that both QM1$'$ and QM2$'$ become successfully entangled with other nodes after $n$ attempts with probability
\begin{equation}
    \mathrm{Pr}(n, P_{\mathrm{ent}}) = [1-(1-P_{\rm ent})^n]^2.
\end{equation}
The average storage time in this case then depends on the average number of attempts needed before entanglement is successfully created and on the time $T_{\rm trial}$ each attempt takes:
\begin{equation}
\langle T_{\rm store}^{\rm case 4}\rangle = T_{\rm trial}\sum_{n=0}^{\infty} n [\mathrm{Pr}(n,P_{\rm ent})- \mathrm{Pr}(n-1,P_{\rm ent})] = \frac{3-2P_{\mathrm{ent}}}{2- P_{\mathrm{ent}}}\langle T_{\rm ent}\rangle.
\end{equation}
This implies that the overall average storage time is
\begin{equation}
    \langle T_{\rm store}\rangle=\frac{\langle T_{\rm store}^{\rm case 1}\rangle+\langle T_{\rm store}^{\rm case 2}\rangle+\langle T_{\rm store}^{\rm case 3}\rangle+\langle T_{\rm store}^{\rm case 4}\rangle}{4}=\frac{7/4-P_{\rm ent}}{2-P_{\rm ent}}\langle T_{\rm ent}\rangle.
\end{equation}
The upper bound on the rate per matter qubit is therefore
\begin{equation}
    R_{\mathrm{max}}^{(2QM)}=\frac{1}{(\langle T_{\rm ent}\rangle+\langle T_{\rm store}\rangle)N_m}=\frac{2-P_{\rm ent}}{15-8P_{\rm ent}}\frac{c}{L},
\end{equation}
where we set $\langle T_{\rm ent}\rangle\to2L_0/c$ and $N_m=2L/L_0$ as appropriate for the memory-based scheme. If we set $P_{\rm ent} \rightarrow 1$~\cite{Hu2008a}, this becomes
\begin{equation}
    R_{\mathrm{max}}^{(2QM)}=\frac{c}{7L}\quad\hbox{for}\quad P_{\rm ent}=1.
\end{equation}

\section*{Classical signaling does not limit the repetition rate of the RGS protocol}

In this section, we show that classical signaling does not limit the repetition rate of the RGS protocol. This is important to consider because some of the $X$ and $Z$ measurements that are performed on the RGS will, depending on the measurement outcomes, leave the remaining photons in a state that is only locally equivalent to a graph state. In this case, additional single-qubit gates must be applied to bring the state to a proper graph state before the next measurements are performed. Otherwise, the bases for these subsequent measurements will be effectively rotated, and the measurements will not modify the state in the intended way. In the following, we show that the extra single-qubit gates only need to be applied on qubits located at the same measurement node where the measurements on which they depend occur. Thus, classical communication between nodes is not needed. The only exception to this occurs at the very end of the protocol, when we are left with a linear cluster state in which the two end qubits are those of Alice and Bob. The final step is to turn this state into a Bell pair shared by Alice and Bob by performing $X$ measurements on the inner qubits of the linear cluster state. Afterward, Alice and Bob's qubits form a Bell pair whose Pauli frame depends on the $X$ measurement outcomes. This information therefore needs to be transferred to them using classical signaling.  However, the protocol can be repeated before this classical signal is received and thus, the classical signaling does not limit the repetition rate. The latter is only limited by the generation time of RGSs.

\subsection*{Recursive construction of a repeater graph state.}

\begin{figure*}
    \centering
    \includegraphics[width=14cm]{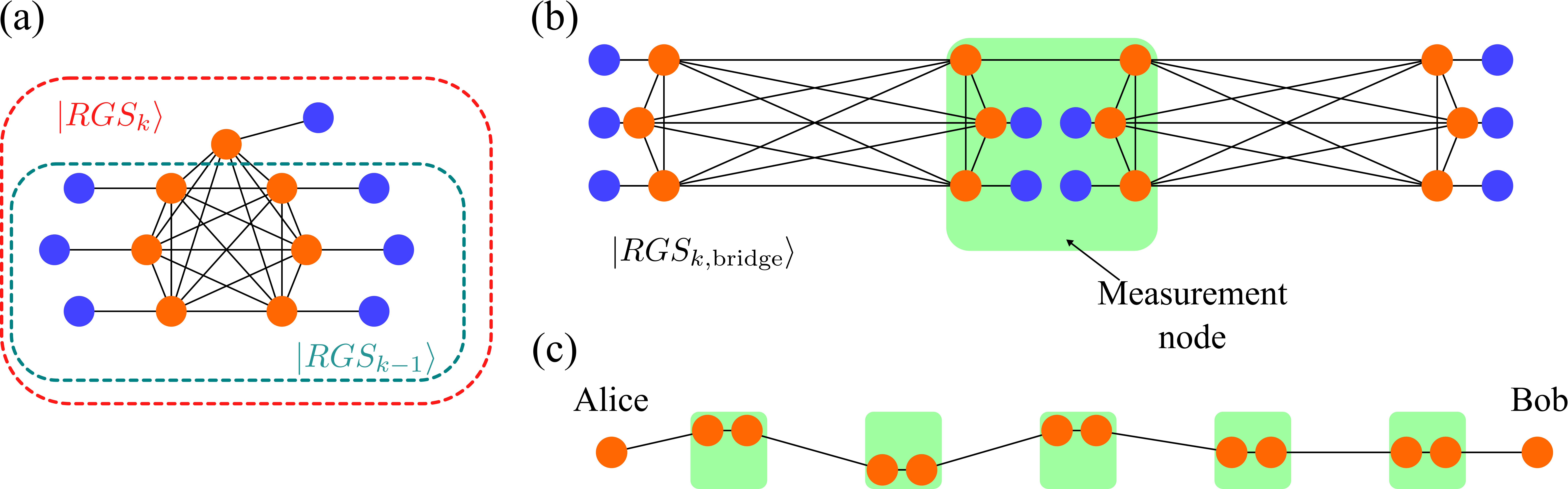}
    \caption{(a) Recursive definition of the quantum state $\ket{RGS_k}$. (b) Realization of a bridge between two RGSs $\ket{RGS_{k,\mathrm{bridge}}}$. (c) Linear cluster state shared between Alice, Bob and all the measurement nodes.
    }
    \label{fig_classical_signaling}
\end{figure*}

We denote the RGS with $k$ arms by $\ket{RGS_k}$. For the purposes of this argument, we allow $k$ to be either odd or even. We want to define this quantum state recursively. As illustrated in Fig.~\ref{fig_classical_signaling}(a), we can make this state by starting with an RGS with $k-1$ arms and then adding one new 1\st-leaf qubit and one new 2\nd-leaf qubit by applying $CZ$ gates between these two new qubits and between the new 1\st-leaf qubit and all the existing 1\st-leaf qubits in $\ket{RGS_{k-1}}$.
In other words, we can start from a 2-qubit cluster state and $\ket{RGS_{k-1}}$ in a product state:
$(\ket{0_{k_1},+_{k_2}} + \ket{1_{k_1},-_{k_2}}) \otimes \ket{RGS_{k-1}}$.
Here, $k_1$ and $k_2$ denote the 1\st-leaf qubit and the 2\nd-leaf qubit from the k\nth arm. We ignore the normalization factors troughout this section. The remaining step is to apply $CZ$ gates between $k_1$ and all the 1\st-leaf qubits in $\ket{RGS_{k-1}}$, which yields
\begin{equation}
    \ket{RGS_k} = \ket{0_{k_1},+_{k_2}} \otimes \ket{RGS_{k-1}} + \ket{1_{k_1},-_{k_2}} \otimes \ket{\overline{RGS}_{k-1}},\label{eq:RGSdecomp}
\end{equation}
where we have introduced the notation
\begin{equation}
    \ket{\overline{RGS}_{k}} \equiv \prod_{i=1}^{k} Z_{i_1} \ket{RGS_{k}}.
\end{equation}
Here, $Z_{i_j}$ is the Pauli $Z$ operator applied to the j\nth-leaf qubit of the i\nth arm of the RGS.

Note that if the RGS has a logical tree encoding, then instead of attaching a 2-qubit cluster state to $\ket{RGS_{k-1}}$, we need to attach a star graph with $b_0$ arms (in the case of a depth-two tree), where $b_0$ is the first branching parameter of the tree. To create $\ket{RGS_{k}}$, each arm of the star graph needs to connect to every 1\st-leaf qubit of $\ket{RGS_{k-1}}$; this requires a total of $b_0(k-1)$ $CZ$ gates. The resulting RGS can still be decomposed as in Eq.~\eqref{eq:RGSdecomp}, but we need to replace $\ket{0_{k_1}}$ by a sum over all the even-parity bit strings on $b_0$ qubits, and $\ket{1_{k_1}}$ needs to be replaced by all the odd-parity bit strings on $b_0$ qubits. In what follows, we will continue to use Eq.~\eqref{eq:RGSdecomp} with the understanding that the $k_1$ qubit states need to be replaced in this manner if a logical tree encoding is included. This does not otherwise alter the analysis below.

\subsection*{Bell state measurement}

Let's now consider the effect of a Bell measurement on 2\nd-leaf qubits from two different RGSs. We suppose that one of these qubits resides in arm $k$ of one RGS, and the other in arm $k'$ of the other RGS. To analyze this measurement, we expand the two-RGS product state using the recursive definition of an RGS shown above:
\begin{equation}
    \begin{aligned}
        \ket{RGS_k} \otimes \ket{RGS'_k} & =  \ket{RGS_{k-1}} \otimes \ket{0_{k_1}, +_{k_2}, +_{{k'}_2}, 0_{{k'}_1}} \otimes \ket{RGS'_{k-1}} \\
        & + \ket{RGS_{k-1}} \otimes \ket{0_{k_1}, +_{k_2}, -_{{k'}_2}, 1_{{k'}_1}} \otimes \ket{\overline{RGS'}_{k-1}}  \\
        & +  \ket{\overline{RGS}_{k-1}} \otimes \ket{1_{k_1}, -_{k_2}, +_{{k'}_2}, 0_{{k'}_1}} \otimes \ket{RGS'_{k-1}} \\
        & + \ket{\overline{RGS}_{k-1}} \otimes \ket{1_{k_1}, -_{k_2}, -_{{k'}_2}, 1_{{k'}_1}} \otimes \ket{\overline{RGS'}_{k-1}}.
    \end{aligned}
\end{equation}
We want to perform a Bell state measurement on qubits $k_2$ and ${k'}_2$. After the Bell measurement, we want to obtain the state:
\begin{equation}
    \begin{aligned}
        \ket{RGS_{k,\mathrm{bridge}}} & =  \ket{RGS_{k-1}} \otimes \ket{0_{k_1}, 0_{{k'}_1}} \otimes \ket{RGS'_{k-1}} \\
        & + \ket{RGS_{k-1}} \otimes \ket{0_{k_1}, 1_{{k'}_1}} \otimes \ket{\overline{RGS'}_{k-1}}  \\
        & +  \ket{\overline{RGS}_{k-1}} \otimes \ket{1_{k_1}, 0_{{k'}_1}} \otimes \ket{RGS'_{k-1}} \\
        & - \ket{\overline{RGS}_{k-1}} \otimes \ket{1_{k_1}, 1_{{k'}_1}} \otimes \ket{\overline{RGS'}_{k-1}},
    \end{aligned}
\end{equation}
which corresponds to the state represented in Fig.~\ref{fig_classical_signaling}(b) where an edge replaces a pair of 2\nd-leaf qubits shared by two RGSs.

The Bell state measurements are realized in the basis:
\begin{equation}
    \begin{aligned}
        \ket{\phi^{\pm}} = \ket{0,+}\pm \ket{1,-}, \\
        \ket{\psi^{\pm}} = \ket{1,+} \pm \ket{0,-}.
    \end{aligned}
\end{equation}
These are not the standard Bell states but are related to them by a Hadamard gate on the second qubit. By applying such a Hadamard gate, a standard Bell state analyzer can be used.
Using this basis for the 2\nd-leaf qubits $k_2$ and ${k'}_2$, we can rewrite the state before the Bell measurement as
\begin{equation}
    \begin{aligned}
        \ket{RGS_k} \otimes \ket{RGS'_k} & =  \ket{RGS_{k-1}} \otimes \ket{0}_{k_1} \otimes \left(\ket{\phi^+}+\ket{\phi^-}+\ket{\psi^+}+\ket{\psi^-}\right)_{k_2, {k'}_2} \otimes \ket{0}_{{k'_1}} \otimes \ket{RGS'_{k-1}} \\
        & + \ket{RGS_{k-1}} \otimes \ket{0}_{k_1} \otimes \left(\ket{\phi^+}-\ket{\phi^-}+\ket{\psi^+}-\ket{\psi^-}\right)_{k_2, {k'}_2}  \otimes \ket{1}_{{k'_1}} \otimes \ket{\overline{RGS'}_{k-1}}  \\
        & +  \ket{\overline{RGS}_{k-1}} \otimes \ket{1}_{k_1} \otimes \left(\ket{\phi^+}+\ket{\phi^-}-\ket{\psi^+}-\ket{\psi^-}\right)_{k_2, {k'}_2}  \otimes \ket{0}_{{k'_1}} \otimes \ket{RGS'_{k-1}} \\
        & + \ket{\overline{RGS}_{k-1}} \otimes \ket{1}_{k_1} \otimes \left(-\ket{\phi^+}+\ket{\phi^-}+\ket{\psi^+}-\ket{\psi^-}\right)_{k_2, {k'}_2}  \otimes \ket{1}_{{k'_1}} \otimes \ket{\overline{RGS'}_{k-1}}.
    \end{aligned}
\end{equation}

Suppose that we can resolve only the two states $\ket{\psi^{\pm}}$ with a linear optics Bell state analyzer. Depending on the measurement outcomes, the state becomes:
\begin{equation}
    \begin{aligned}
        \ket{\psi^-} &\rightarrow Z_{k_1} Z_{{k'}_1}\ket{RGS_{k,\mathrm{bridge}}}, \\
        \ket{\psi^+} &\rightarrow Z_{k_1} \ket{RGS_{k,\mathrm{bridge}}}.
    \end{aligned}
\end{equation}

Thus, we find that if the outcome is $\ket{\psi^+}$ (respectively $\ket{\psi^-}$) we need to apply a $Z$ rotation on qubit $k_1$ (on qubits $k_1$ and ${k'}_1$) to recover the desired state. Note that this qubit is at the measurement node where we perform the Bell measurement.

\subsection*{$Z$ measurements}

If the measurement fails or if one Bell measurement has already succeeded, we need to perform $Z$ measurements on 1\st-leaf qubits to disconnect them and their associated 2\nd-leaf qubits from the graph. Depending on the outcomes of these measurements, the resulting state is one of four possible states:
\begin{equation}
    \begin{aligned}
        \ket{RGS_{k-1}}\otimes \ket{RGS'_{k-1}}, \\
        \ket{\overline{RGS}_{k-1}}\otimes \ket{RGS'_{k-1}}, \\
        \ket{RGS_{k-1}}\otimes \ket{\overline{RGS'}_{k-1}}, \\
        \ket{\overline{RGS}_{k-1}}\otimes \ket{\overline{RGS'}_{k-1}}.
    \end{aligned}
\end{equation}
Note that if a Bell measurement has already succeeded, then each of the above outcomes needs to be acted upon by a $CZ$ gate on the two 1\st-leaf qubits associated with the successful measurement. We leave this possible $CZ$ gate implicit in the following.
Ideally, we want to recover $\ket{RGS_{k-1}}\otimes \ket{RGS'_{k-1}}$. Therefore, we need to be able to convert a state $\ket{\overline{RGS}_{k-1}}$ into $\ket{RGS_{k-1}}$ by performing measurements only on qubits present in the measurement node.
Let's consider any other arm $l$ of the RGS that is present at the same measurement node and rewrite the state as
\begin{equation}
     \ket{RGS_{k-1}} = \ket{0_{l_1}, +_{l_2}} \otimes \ket{RGS_{k-2}} + \ket{1_{l_1}, -_{l_2}} \otimes \ket{\overline{RGS}_{k-2}}.
\end{equation}
By applying $Z$ gates on all the 1\st-leaf qubits, we also have:
\begin{equation}
     \ket{\overline{RGS}_{k-1}} = \ket{0_{l_1}, +_{l_2}} \otimes \ket{\overline{RGS}_{k-2}} - \ket{1_{l_1}, -_{l_2}}) \otimes \ket{RGS_{k-2}}.
\end{equation}
A comparison of these two results leads to the identity:
\begin{equation}
    \begin{aligned}
        Y_{l_1}Z_{l_2}\ket{\overline{RGS}_{k-1}} = i \ket{RGS_{k-1}}.
    \end{aligned}
\end{equation}
By applying local rotations of qubits situated at the same measurement node, it is therefore possible to recover the desired state.
The last two $Z$ measurements on the two qubits denoted $k_1$ and $k'_1$ requires special care as the only two other qubits, $l_1$ and $l'_1$, positioned at this measurement node, are the one which connect the two RGSs (if a Bell measurement has succeeded) and should undergo an $X$ measurement. In that case, it is still possible to use $l_1$ and $l'_1$ to recover the desired graph state by performing single-qubit rotations that depends on the measurement outcomes of $k_1$ and $k'_1$:
    \begin{equation}
        \begin{aligned}
        \ket{0_{k_1},0_{k'_1}} & \rightarrow I_{l_1} \otimes I_{l'_1}, \\
        \ket{0_{k_1},1_{k'_1}} & \rightarrow - i Z_{l_1} \otimes Y_{l'_1}, \\
        \ket{1_{k_1},0_{k'_1}} & \rightarrow - i Y_{l_1} \otimes Z_{l'_1}, \\
        \ket{1_{k_1},1_{k'_1}} & \rightarrow - X_{l_1} \otimes X_{l'_1}, \\
        \end{aligned}
    \end{equation}
    where $I_i$ denotes the identity on qubit $i$.

\subsection*{$X$ measurements and classical signaling}

If at least one Bell measurement succeeds at each measurement node and if all the required $Z$ measurements succeed, we are left with a linear cluster state with two 1\st-leaf qubits at each measurement node (see Fig.~\ref{fig_classical_signaling}(c)).
The next step is to perform $X$ measurements on these qubits. Performing $X$ measurements on two connected qubits of a graph state has the effect of connecting each qubit in the neighborhood of the first qubit to all the qubits in the neighborhood of the second one~\cite{Varnava2006}. Therefore, after the $X$ measurements of all the qubits at each measurement node, the two remaining qubits shared by Alice and Bob are in a maximally entangled state. More precisely, this state is equivalent to a Bell pair up to local qubit rotations. It is important to note that all the $X$ measurements can be performed independently of each other. Although the state can deviate from a proper graph state depending on the measurement outcomes, it is straightforward to show that the state only differs by at most local $Z$ gates. This can effectively convert some of the $X$ measurements to $Y$ measurements, but both types of measurements have a similar effect on the state, and a Bell pair is still obtained once all the measurements are completed. However, the precise state depends on all the measurement outcomes of each $X$ measurement. This information needs to be sent via a classical channel to Alice and Bob to recover the desired state.

This classical signal does not limit the repetition rate of the protocol, which can either be used for quantum teleportation of quantum states or for QKD. Indeed, for quantum teleportation, two quantum memories situated at Alice's and Bob's nodes can be used to store the Bell pair while waiting for the classical signal to arrive. In that case, the protocol can still be repeated at a high rate as long as there are enough quantum memories at Alice's and Bob's nodes to store the states that are already transferred. It is still the case that the protocol does not require quantum memories at the nodes inside the network, which is a clear resource improvement compared to the memory-based approach.
For QKD, Alice's and Bob's qubits can be measured in a random basis upon arrival at their nodes, and they don't need to be stored in quantum memories until the classical signaling is received. The classical signaling is still required to recover the Pauli frame of the measured qubits and to check if Alice's and Bob's qubits were measured in compatible basis for QKD.

    \section*{Error correction using tree graph states}

    In the main text, we have shown that logical encoding with tree graph states allows measuring a qubit with a probability $\mathrm{Pr}(M_{Z,\ell})$ or $\mathrm{Pr}(M_{X,\ell})$ that is higher than the single-photon measurement probability $P_{\rm ph}$. We now show that the logical encoding based on tree graphs also increases the tolerance of the RGS protocol to single-qubit errors.
    Indeed, if $m_k$ indirect measurements are performed on a qubit at level $k$ in the tree, it is possible to use a majority vote of the measurements to reduce the effect of errors. Suppose that all $b_k$ of the possible indirect measurements are performed, where $b_k$ is the $k$\nth branching parameter of the tree.
    The probability that exactly $m_k$ of these measurements succeed (perhaps with errors) is
    \begin{equation}
        p_k(m_k) = \binom{b_k}{m_k} {s_k}^{m_k} {(1 - s_k)}^{b_k - m_k},
    \end{equation}
    where $s_k$ is the probability that a single indirect measurement succeeds.

    The average error after $b_k$ indirect measurement attempts at level k is:
    \begin{equation}
        e_{I_k} = \frac{1}{r_k} \sum_{m_k=1}^{b_k} p_k(m_k) e_{I_k|m_k},
    \end{equation}
    where $e_{I_k|m_k}$ is the average error in the case of $m_k$ successful measurements, and $r_k$ is the probability that at least one indirect measurement succeeds. For $m_k$ indirect measurements, an error still occurs in the majority vote if more than half of the indirect measurements ($m_k / 2$) are faulty. Thus, if $m_k$ is odd:
    \begin{equation}
        \begin{aligned}
            e_{I_k|m_k} &=\mathrm{Pr}(N_{\mathrm{errors}} > m_k/2) \\
            &= \sum_{j=\lceil m_k/2 \rceil}^{m_k} \binom{m_k}{j} {(e_{I_k|1})}^j {(1 - e_{I_k|1})}^{m_k - j},
        \end{aligned}
    \end{equation}
    where $e_{I_k|1}$ is the average error of a single indirect measurement. If $m_k$ is even, we have:
    \begin{equation}
            e_{I_k|m_k} = \sum_{j=\lceil m_k/2 \rceil}^{m_k - 1} \binom{m_k - 1}{j} {(e_{I_k|1})}^j {(1 - e_{I_k|1})}^{m_k -1 - j}.
    \end{equation}
    The sum goes only up to $m_k-1$ because we can randomly remove one result to return to the odd-number case and re-use the above formula for the probability.

    The single indirect $Z$ measurement error $e_{I_k|1}$ of a qubit A at level $k$ depends on the error of the $X$ measurement of qubit B at level $k+1$ (which we take to be a constant error probability $\epsilon$) and on the errors of the $b_{k+1}$ measurements on the  qubits at level $k+2$, denoted $C_i$ in Fig.~\ref{fig_generation}(a). These can either be direct (with error $\epsilon$) or indirect (with error $e_{I_{k+2}}$). Let $n_k$ be the number of qubits that are only measured directly (with error $\epsilon$) while $b_{k+1} - n_k$ qubits are measured indirectly (with error $e_{I_{k+2}}$). $e_{I_k|1}$ is thus given by:
    \begin{equation}
        e_{I_k|1} = \sum_{n_k = 0}^{b_{k+1}} \binom{b_{k+1}}{n_k} \mathrm{Pr}(I_{Z,k+2}|M_{Z,k+2})^{b_{k+1}-n_k} (1 - \mathrm{Pr}(I_{Z,k+2}|M_{Z,k+2}))^{n_k} e_{n_k},
    \end{equation}
    where $e_{n_k}$ is the error in the case of $n_k$ direct qubit measurements, and
    \begin{equation}
        \mathrm{Pr}(I_{Z, k}|M_{Z,k}) = \frac{\mathrm{Pr}(I_{Z,k}) \mathrm{Pr}(M_{Z,k}|I_{Z,k})}{\mathrm{Pr}(M_{Z,k})} = \frac{r_k}{\mathrm{Pr}(M_{Z,k})}.
    \end{equation}
    Here, $I_{Z,k}$ denotes a successful indirect $Z$ measurement of a photon at level $k$ and thus $\mathrm{Pr}(I_{Z,k})$ is what is called $r_k$ in the main text.  Therefore, $\mathrm{Pr}(I_{Z,k}|M_{Z,k})$ is the probability that a photon at level $k$ was successfully measured indirectly ($I_{Z,k}$) given that it has been successfully measured ($M_{Z,k}$).
    Note that a photon can be both successfully directly and indirectly measured. In that case, we keep the indirect measurement outcome which should have a smaller error $e_{I_k}$ compared to the intrinsic single qubit error $\epsilon$, thanks to the error correction.

    The only remaining quantity to determine is the error $e_{n_k}$. The $X$ measurement on qubit B and the $n_k$ direct $Z$ measurements on the qubits $C_i$ are performed with error $\epsilon$, while the remaining $b_{k+1}-n_k$ indirect $Z$ measurements are performed with error $e_{I_{k+2}}$. Note that the indirect measurement outcome is given by the net parity of all the $X$ and $Z$ measurement outcomes. Thus, the indirect measurement can still yield the correct outcome even if there are errors in the $X$ and $Z$ measurements, provided there are an even number of errors so that the parity remains unchanged. Only an odd number of errors causes the indirect measurement to fail.
    Thus:

    \begin{equation}
        e_{n_k} = \sum_{i = 0}^{n_k+1}{\left( \binom{n_k+1}{i}{\epsilon}^i (1- \epsilon)^{n_k+1 - i} \sum_{\substack{j = 0,\\ i+j=1[2]}}^{b_{k+1} - n_k}  \binom{b_{k+1} - n_k}{j}  {e_{I_{k+2}}}^j (1- e_{I_{k+2}})^{b_{k+1} -n_k - j}\right) }
    \end{equation}

    In this equation, we take into account the fact that only odd numbers of measurement errors contribute to $e_{n_k}$ by restricting the second sum to only values of $i+j=1[2]$.
    This is different than what was obtained in Ref.~\cite{Azuma2015}, perhaps due to additional approximations in that reference.

        \begin{figure*}[!ht]
        \centering
        \includegraphics[width=14cm]{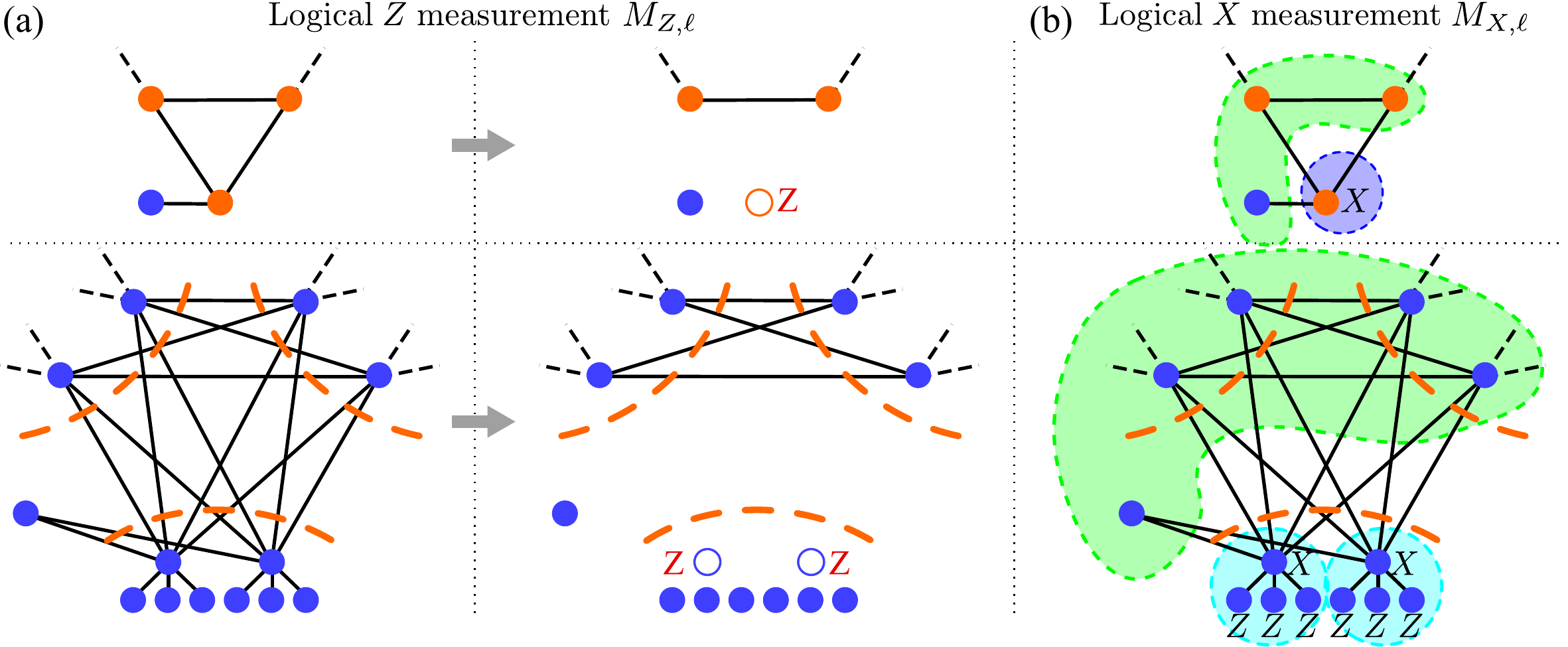}
        \caption{(a) Logical $Z$ measurement $M_{Z,\ell}$ at the physical level. (b) Logical $X$ measurement $M_{X,\ell}$ at the physical level. The parity of the $Z$ measurements on the qubits in the green area is the same as the logical $M_{X,\ell}$ measurement on the logical qubit (purple) and is the same as the parity measurement on the two light blue areas (where measurement bases are indicated next to each qubit).
        }
        \label{fig_logical_measurement}
    \end{figure*}

    As shown in Fig.~\ref{fig_logical_measurement}(a), the logical $Z$ measurement $M_{Z,\ell}$
    succeeds  if all the 1\st-level qubits are measured directly or indirectly in the $Z$ basis. The logical $X$ measurement $M_{X,\ell}$ succeeds if any of the 1\st level qubits is measured in the $X$ basis and all its corresponding 2\nd level qubits are successfully measured in the $Z$ basis either directly or indirectly, as shown in Ref.~\cite{Azuma2015}.
    To understand this, let's denote by $L$ the logical qubit that is connected to a set $A = \{A_1, A_2, ...\}$ of other qubits, represented by the green area in Fig.~\ref{fig_logical_measurement}(b).
    Because $X_L Z_A = X_L \otimes_i Z_{A_i}$ is a stabilizer of the RGS at the logical level, the logical $X$ measurement of qubit $L$, $M_{X,\ell}$, has the same outcome as $Z_A$, i.e. the parity of the product of all $Z$ measurements on the qubits in $A$. At the physical level, the logical qubit is encoded with a tree graph state where we consider a first-level qubit $B$ and its set of neighbor qubits in the tree $C(B)$.
    $X_B Z_{C(B)} Z_A$ is a stabilizer of the RGS at the physical level and thus $X_B Z_{C(B)}$ (represented by either of the light blue areas in Fig.~\ref{fig_logical_measurement}(b)) has the same parity as $Z_A$ and therefore constitutes an $X$ measurement of the logical qubit $L$.

    The logical $X$ and $Z$ errors are thus given by:
    \begin{equation}
        \begin{aligned}
            \bar{e}_X & = e_{I_0},  \\
            \bar{e}_Z & = \sum_{n = 0}^{b_{0}} \binom{b_{0}}{n} \mathrm{Pr}(I_{1}|Z_{1})^{b_{0}-n} (1 - \mathrm{Pr}(I_{1}|Z_{1}))^{n} e_{n}, \\
            \mathrm{with} \; e_{n} & = \sum_{i = 0}^{n}{\left( \binom{n}{i}{\epsilon}^i (1- \epsilon)^{n - i} \sum_{\substack{j = 0,\\ i+j=1[2]}}^{b_{0} - n}  \binom{b_{0} - n}{j}  {e_{I_{1}}}^j (1-  e_{I_{1}})^{b_{0} -n - j}\right). }
        \end{aligned}
    \end{equation}
    The derivation of $\bar{e}_Z$ follows the same idea as $e_{I_{k}|1}$ (for "$k=-1$") except that no $X$ measurements are needed here.

        \subsection*{Fidelity of the full protocol}

        We have used the results already derived in the supplementary materials of Ref.~\cite{Azuma2015} to calculate the global fidelity of the generated entangled pair of photons shared by Alice and Bob. The total fidelity $F_{AB}$ is given by
        \begin{equation}
            F_{AB} = 1 - (E_x + E_y + E_z),
        \end{equation}
        with
        \begin{equation}
            \begin{aligned}
                E_x & = E_z = \frac{1}{4} \left(1 - (1 - 2 \epsilon)^{2(N_{QR} + 1)} (1 - 2 \bar{e}_X)^{2N_{QR}} \right),\\
                E_y & = \frac{1}{4} \left( 1 + (1 - 2 \epsilon)^{2(N_{QR} +1)} (1 - 2 \bar{e}_X)^{2 N_{QR}} -2 (1 - 2 \epsilon)^{2 (N_{QR} + 1)} (1 - 2 \bar{e}_X)^{N_{QR}} (1 - 2 \bar{e}_Z)^{(2m - 2) N_{QR}}   \right).
            \end{aligned}
        \end{equation}

\end{document}